\documentclass[twocolumn,psfig,aps,showpacs]{revtex4}
\usepackage{amsfonts}
\usepackage{amsmath}
\usepackage{graphicx}
\usepackage{bm}
\usepackage{amssymb}
\usepackage{dcolumn}

\begin{document}

\title{Magnetism and Thermodynamics of Spin-$1/2$ Heisenberg Diamond Chains
in a Magnetic Field}
\author{Bo Gu and Gang Su$^{\ast }$}
\affiliation{College of Physical Sciences, Graduate University of Chinese Academy of
Sciences, P. O. Box 4588, Beijing 100049, China}

\begin{abstract}
The magnetic and thermodynamic properties of spin-$1/2$ Heisenberg diamond
chains are investigated in three different cases: (a) $J_{1}$, $J_{2}$, $%
J_{3}>0$ (frustrated); (b) $J_{1}$, $J_{3}<0$, $J_{2}>0$ (frustrated); and
(c) $J_{1}$, $J_{2}>0$, $J_{3}<0$ (non-frustrated), where the bond coupling $%
J_{i}$ $(i=1,2,3)>0$ stands for an antiferromagnetic (AF) interaction, and $%
<0$ for a ferromagnetic (F) interaction. The density matrix renormalization
group (DMRG) technique is invoked to study the properties of the system in
the ground state, while the transfer matrix renormalization group (TMRG)
technique is applied to explore the thermodynamic properties. The local
magnetic moments, spin correlation functions, and static structure factors
are discussed in the ground state for the three cases. It is shown that the
static structure factor $S(q)$ shows peaks at wavevectors $q=a\pi /3$ $%
(a=0,1,2,3,4,5)$ for different couplings in a zero magnetic field, which,
however in the magnetic fields where the magnetization plateau with $m=1/6$
pertains, exhibits the peaks only at $q=0$, $2\pi /3$ and $4\pi /3$, which
are found to be couplings-independent. The DMRG results of the zero-field
static structure factor can be nicely fitted by a linear superposition of
six modes, where two fitting equations are proposed. It is observed that the
six modes are closely related to the low-lying excitations of the system. At
finite temperatures, the magnetization, susceptibility and specific heat
show various behaviors for different couplings. The double-peak structures
of the susceptibility and specific heat against temperature are obtained,
where the peak positions and heights are found to depend on the competition
of the couplings. It is also uncovered that the XXZ anisotropy of F and AF
couplings leads the system of case (c) to display quite different behaviors.
In addition, the experimental data of the susceptibility, specific heat and
magnetization for the compound Cu$_{3}$(CO$_{3}$)$_{2}$(OH)$_{2}$ are fairly
compared with our TMRG results.
\end{abstract}

\maketitle

\section{INTRODUCTION}

Low-dimensional quantum spin systems with competing interactions have become
an intriguing subject in the last decades. Among many achievements in this
area, the phenomenon of the topological quantization of magnetization has
attracted much attention both theoretically and experimentally. A general
necessary condition for the appearance of the magnetization plateaus has
been proposed by Oshikawa, Yamanaka and Affleck (OYA) \cite{OYA}, stating
that for the Heisenberg antiferromagnetic (AF) spin chain with a single-ion
anisotropy, the magnetization curve may have plateaus at which the
magnetization per site $m$ is topologically quantized by $n(S-m)=integer$,
where $S$ is the spin, and $n$ is the period of the ground state determined
by the explicit spatial structure of the Hamiltonian. As one of fascinating
models which potentially possesses the magnetization plateaus, the
Heisenberg diamond chain, consisting of diamond-shaped topological unit
along the chain, as shown in Fig. \ref{chain}, has also gained much
attention both experimentally and theoretically (e.g. Refs. \cite%
{Drillon,Sakurai,Ishii,Fujisawa,CuCOOH1,CuCOOH2,TKS,aaa1,aaa2,aaa3,aaa4,aaa5,faf1,faf2}%
).

It has been observed that the compounds, A$_{3}$Cu$_{3}$(PO$_{4}$)$_{4}$
with A = Ca, Sr\cite{Drillon}, and Bi$_{4}$Cu$_{3}$V$_{2}$O$_{14}$\cite%
{Sakurai} can be nicely modeled by the Heisenberg diamond chain. Another
spin-$1/2$ compound Cu$_{3}$Cl$_{6}$(H$_{2}$O)$_{2}$$\cdot $2H$_{8}$C$_{4}$SO%
$_{2}$ was initially regarded as a model substance for the spin-$1/2$
diamond chain \cite{Ishii}, but a later experimental research reveals that
this compound should be described by a double chain model with very weak
bond alternations, and the lattice of the compound is found to be Cu$_{2}$Cl$%
_{4}$$\cdot $H$_{8}$C$_{4}$SO$_{2}$ \cite{Fujisawa}. Recently, Kikuchi 
\textit{et al.}\cite{CuCOOH1} have reported the experimental results on a
spin-$1/2$ compound Cu$_{3}$(CO$_{3}$)$_{2}$(OH)$_{2}$, where local Cu$^{2+}$
ions with spin $S=1/2$ are arranged along the chain direction, and the
diamond-shaped units consist of a one-dimensional (1D) lattice. The $1/3$
magnetization plateau and the double peaks in the magnetic susceptibility as
well as the specific heat as functions of temperature have been observed
experimentally\cite{CuCOOH1, CuCOOH2}, which has been discussed in terms of
the spin-$1/2$ Heisenberg diamond chain with AF couplings $J_{1}$, $J_{2}$
and $J_{3}>0$.

On the theoretical aspect, the frustrated diamond spin chain with AF
interactions $J_{1}$, $J_{2}$ and $J_{3}>0$ was studied by a few groups. The
first diamond spin chain was explored under a symmetrical condition $%
J_{1}=J_{3}$\cite{TKS}. Owing to the competition of AF interactions, the
phase diagram in the ground state of the spin-$1/2$ frustrated diamond chain
was found to contain different phases, in which the magnetization plateaus
at $m=1/6$ as well as $1/3$ are predicted\cite{aaa1,aaa2,aaa3,aaa4,aaa5}.
Another frustrated diamond chain with ferromagnetic (F) interactions $J_{1}$%
, $J_{3}<0$ and AF interaction $J_{2}>0$ was also investigated
theoretically, which can be experimentally realized if all angles of the
exchange coupling bonds are arranged to be around $90^{\circ }$, a region
where it is usually hard to determine safely the coupling constants and even
about their signs\cite{faf1,faf2}. Despite of these works, the
investigations on the Heisenberg diamond spin chain with various competing
interactions are still sparse.

Motivated by the recent experimental observation on the azurite compound Cu$%
_{3}$(CO$_{3}$)$_{2}$(OH)$_{2}$\cite{CuCOOH1, CuCOOH2}, we shall explore
systematically the magnetic and thermodynamic properties of the spin-$1/2$
Heisenberg diamond chain with various competing interactions in a magnetic
field, and attempt to fit into the experimental observation on the azurite
in a consistent manner. The density matrix renormalization group (DMRG) as
well as the transfer matrix renormalization group (TMRG) techniques will be
invoked to study the ground-state properties and thermodynamics of the model
under interest, respectively. The local magnetic moments, spin correlation
functions, and static structure factors will be discussed for three cases at
zero temperature. It is found that the static structure factor $S(q)$ shows
peaks in zero magnetic field at wavevectors $q=a\pi /3$ $(a=0,1,2,3,4,5)$
for different couplings, while in the magnetic fields where the
magnetization plateau with $m=1/6$ remains, the peaks appear only at
wavevectors $q=0$, $2\pi /3$ and $4\pi /3$, which are found to be
couplings-independent. These information could be useful for further neutron
studies. The double-peak structures of the susceptibility and specific heat
against temperature are obtained, where the peak positions and heights are
found to depend on the competition of the couplings. It is uncovered that
the XXZ anisotropy of F and AF couplings leads the system without
frustration (see below) to display quite different behaviors. In addition,
the experimental data of the susceptibility, specific heat and magnetization
for the compound Cu$_{3}$(CO$_{3}$)$_{2}$(OH)$_{2}$ are fairly compared with
our TMRG results.

The rest of this paper is outlined as follows. In Sec. II, we shall
introduce the model Hamiltonian for the spin-$1/2$ Heisenberg diamond chain
with three couplings $J_{1}$, $J_{2}$ and $J_{3}$, where three particular
cases are identified. In Sec. III, the magnetic and thermodynamic properties
of a frustrated diamond chain with AF interactions $J_{1}$, $J_{2}$ and $%
J_{3}>0$ will be discussed. In Sec. IV, the physical properties of another
frustrated diamond chain with F interactions $J_{1}$, $J_{3}<0$ and AF
interaction $J_{2}>0$ will be considered. In Sec. V, the magnetism and
thermodynamics of a non-frustrated diamond chain with AF interactions $J_{1}$%
, $J_{2}>0$ and F interaction $J_{3}<0$ will be explored, and a comparison
to the experimental data on the azurite compound will be made. Finally, a
brief summary and discussion will be presented in Sec. VI.

\section{MODEL}

\begin{figure}[tbp]
\includegraphics[width = 8.5cm]{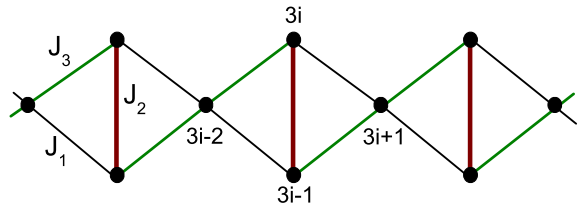}
\caption{ (Color online) Sketch of the Heisenberg diamond chain. The bond
interactions are denoted by $J_{1}$, $J_{2}$, and $J_{3}$. Three cases will
be considered: (a) $J_{1}$, $J_{2}$, $J_{3}>0$ (a frustrated diamond chain);
(b) $J_{1}$, $J_{3}<0$, $J_{2}>0$ (a frustrated diamond chain with competing
interactions); and (c) $J_{1}$, $J_{2}>0$, $J_{3}<0$ (a diamond chain
without frustration). Note that $J_{i}>0$ stands for an antiferromagnetic
interaction while $J_{i}<0$ for a ferromagnetic interaction, where $i=1,2,3$%
. }
\label{chain}
\end{figure}

The Hamiltonian of the spin-$1/2$ Heisenberg diamond chain reads 
\begin{eqnarray}
\mathcal{H} &=&\sum\limits_{i=1}^{L/3}(J_{1}\mathbf{S}_{3i-2}\cdot \mathbf{S}%
_{3i-1}+J_{2}\mathbf{S}_{3i-1}\cdot \mathbf{S}_{3i}  \notag \\
&&+J_{3}\mathbf{S}_{3i-2}\cdot \mathbf{S}_{3i}+J_{3}\mathbf{S}_{3i-1}\cdot 
\mathbf{S}_{3i+1}  \notag \\
&&+J_{1}\mathbf{S}_{3i}\cdot \mathbf{S}_{3i+1})-\mathbf{H}\cdot
\sum\limits_{j=1}^{L}\mathbf{S}_{j},  \label{Ham}
\end{eqnarray}%
where $\mathbf{S}_{j}$ is the spin operator at the $jth$ site, $L$ is the
total number of spins in the diamond chain, $J_{i}$ ($i=1,2,3$) stands for
exchange interactions, $\mathbf{H}$ is the external magnetic field, $g\mu
_{B}=1$ and $k_{B}=1$. $J_{i}>0$ represents the AF coupling while $J_{i}<0$
the F interaction. There are three different cases particularly interesting,
as displayed in Fig. \ref{chain}, which will be considered in the present
paper: (a) a frustrated diamond chain with $J_{1}$, $J_{2}$, $J_{3}>0$; (b)
a frustrated diamond chain with competing interactions $J_{1}$, $J_{3}<0$, $%
J_{2}>0$; and (c) a diamond chain without frustration with $J_{1}$, $J_{2}>0$%
, $J_{3}<0$. It should be remarked that in the case (c) of this model, since
the two end points of the $J_{2}$ bond represent the two different lattice
sites, it is possible that $J_{1}$ and $J_{3}$ can be different, even their
signs.

The magnetic properties and thermodynamics for the aforementioned three spin-%
$1/2$ Heisenberg diamond chains in the ground states and at finite
temperatures will be investigated by means of the DMRG and TMRG methods,
respectively. As the DMRG and TMRG techniques were detailed in two nice
reviews\cite{Review1, Review2}, we shall not repeat the technical details
for concise. In the ground-state calculations, the total number of spins in
the diamond chain is taken at least as $L=120$. At finite temperatures, the
thermodynamic properties presented below are calculated down to temperature $%
T=0.05$ (in units of $|J_{1}|$) in the thermodynamic limit. In our
calculations, the number of kept optimal states is taken as $81 $; the width
of the imaginary time slice is taken as $\varepsilon =0.1$; the
Trotter-Suzuki error is less than $10^{-3}$; and the truncation error is
smaller than $10^{-6}$.

\section{A Frustrated Heisenberg Diamond Chain ($J_{1},J_{2},J_{3}>0$)}

\subsection{Local Magnetic Moment and Spin Correlation Function}

\begin{figure}[tbp]
\includegraphics[width = 8.5cm]{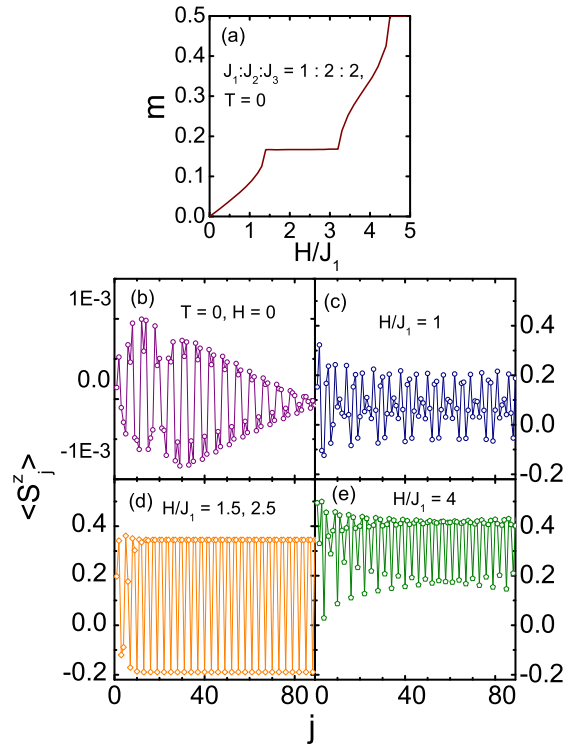}
\caption{ (Color online) For a spin-$1/2$ frustrated Heisenberg diamond
chain with fixed couplings $J_{1}$ : $J_{2}$ : $J_{3}$ = 1 : 2 : 2, (a) the
magnetization per site $m$ as a function of magnetic field $H$ in the ground
states; and the spatial dependence of the averaged local magnetic moment $%
\langle S^{z}_{j} \rangle$ in the ground states with external field (b) $%
H/J_{1}$ = 0, (c) 1, (d) 1.5 and 2.5, and (e) 4.}
\label{aaasj}
\end{figure}

Figure \ref{aaasj}(a) manifests the magnetization process of a frustrated
spin-$1/2$ Heisenberg diamond chain with the couplings satisfying $%
J_{1}:J_{2}:J_{3}=1:2:2$ at zero temperature. The plateau of magnetization
per site $m=1/6$ is observed. According to the OYA necessary condition \cite%
{OYA}, the $m$ = $1/6$ plateau of a spin-$1/2$ Heisenberg chain corresponds
to the period of the ground state $n=3$. Beyond the magnetization plateau
region, the magnetic curve goes up quickly with increasing the magnetic
field $H$. Above the upper critical field, the magnetic curve shows a s-like
shape. To further look at how this magnetization plateau appears, the
spatial dependence of the averaged local magnetic moment $\langle
S_{j}^{z}\rangle $ in the ground states under different external fields is
presented, as shown in Fig. \ref{aaasj}(b). It is seen that in the absence
of external field, the expectation value $\langle S_{j}^{z}\rangle $ changes
its sign at every three sites within a very small range of $%
(-10^{-3},10^{-3})$ because of quantum fluctuations, resulting in the
magnetization per site $m=\sum_{j=1}^{L}\langle S_{j}^{z}\rangle /L=0$. $%
\langle S_{j}^{z}\rangle $ increases with increasing the magnetic field, and
oscillates with increasing $j$, whose unit of three spins is gradually
divided into a pair and a single, as displayed in Fig. \ref{aaasj}(c). At
the field $H/J_{1}=1.5$, as demonstrated in Fig. \ref{aaasj}(d), the
behavior of $\langle S_{j}^{z}\rangle $ falls into a perfect sequence such
as $\{...,(S_{a},S_{a},S_{b}),...\}$ with $S_{a}=0.345$ and $S_{b}=-0.190$,
giving rise to the magnetization per site $m=1/6$. In addition, such a
sequence remains with increasing the magnetic field till $H/J_{1}=2.5$,
implying that the $m=1/6$ plateau appears in the range of $H/J_{1}=1.5\sim
2.5$, as manifested in Fig. \ref{aaasj}(a). When the field is promoted
further, the sequence changes into a waved succession with smaller swing of $%
(S_{a}-S_{b})$, as shown in Fig. \ref{aaasj}(e), which corresponds to the
fact that the plateau state of $m=1/6$ is destroyed, and gives rise to a
s-like shape of $M(H)$. It is noting that when the plateau state of $m=1/6$
is destroyed, the increase of $m$ at first is mainly attributed to a rapid
lift of $S_{b}$, and later, the double $S_{a}$ start to flimsily increase
till $S_{a}=S_{b}=0.5$ at the saturated field.

The physical picture for the above results could be understood as follows.
For the $m=1/6$ plateau state at $J_{1}:J_{2}:J_{3}=1:2:2$, we note that if
an approximate wave function defined by\cite{aaa5} 
\begin{eqnarray}
\psi _{i} &=&\frac{1}{\sqrt{6}}(2|\uparrow _{3i-2}\uparrow _{3i-1}\downarrow
_{3i}\rangle \pm |\uparrow _{3i-2}\downarrow _{3i-1}\uparrow _{3i}\rangle 
\notag \\
&\pm &|\downarrow _{3i-2}\uparrow _{3i-1}\uparrow _{3i}\rangle ),\text{ }%
(i=1,...,L/3)  \label{aaawave}
\end{eqnarray}%
where $\uparrow _{j}(\downarrow _{j})$ denotes spin up (down) on site $j$,
is applied, one may obtain $\langle \psi _{i}|S_{3i-2}^{z}|\psi _{i}\rangle
=1/3$, $\langle \psi _{i}|S_{3i-1}^{z}|\psi _{i}\rangle =1/3$, $\langle \psi
_{i}|S_{3i}^{z}|\psi _{i}\rangle =-1/6$, giving rise to a sequence \{..., ($%
\frac{1}{3}$, $\frac{1}{3}$, $-\frac{1}{6}$), ...\}, and $m=(\frac{1}{3}+%
\frac{1}{3}-\frac{1}{6})/3=1/6$, which is in agreement with our DMRG results
\{..., ($0.345$, $0.345$, $-0.190$), ...\}. This observation shows that the
ground state of this plateau state might be described by trimerized states.

\begin{figure}[tbp]
\includegraphics[width = 8.5cm]{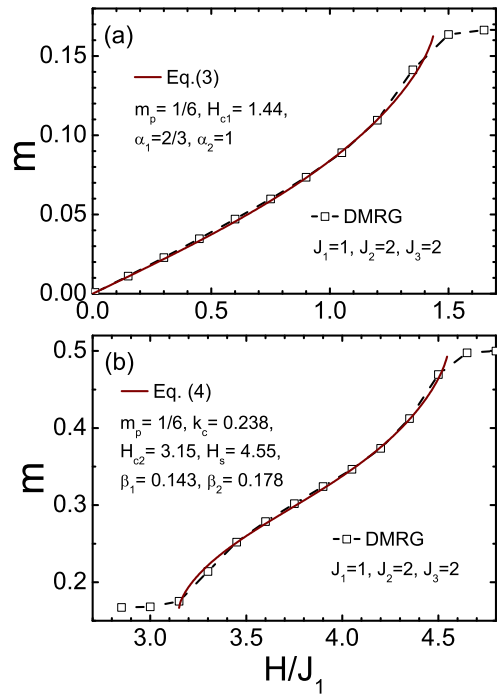}
\caption{ (Color online) For a spin-$1/2$ frustrated Heisenberg diamond
chain with fixed couplings $J_{1}:J_{2}:J_{3}=1:2:2$, the DMRG results of
the magnetization per site $m$ as a function of magnetic field $H$ away from
the plateau state can be fairly fitted by Eqs. (\protect\ref{eqmh-1}) and (%
\protect\ref{eqmh-2}), (a) for $0\leq H\leq H_{c_{1}}$, and (b) for $%
H_{c_{2}}\leq H\leq H_{s}$.}
\label{aaaeqmh}
\end{figure}

Let $H_{c_{1}}$ and $H_{c2}$ be the lower and upper critical magnetic field
at which the magnetization plateau appears and is destructed, respectively.
For $H_{c_{1}}\leq H\leq H_{c2}$, the magnetization $m=m_{p}=1/6$, namely,
the system falls into the magnetization plateau state. For $0\leq H\leq
H_{c_{1}}$ and $J_{1}:J_{2}:J_{3}=1:2:2$, the magnetization curve shows the
following behavior%
\begin{equation}
m(H)=m_{p}(\frac{H}{H_{c_{1}}})[1+\alpha _{1}(1-\frac{H}{H_{c_{1}}})-\alpha
_{2}(1-\frac{H}{H_{c_{1}}})^{2/3}],  \label{eqmh-1}
\end{equation}%
where $H_{c_{1}}/J_{1}=1.44$, the parameters $\alpha _{1}=2/3$ and $\alpha
_{2}=1$. Obviously, when $H=0$, $m=0$; $H=H_{c_{1}}$, $m=m_{p}$. A fair
comparison of Eq. (\ref{eqmh-1}) to the DMRG results is presented in Fig. %
\ref{aaaeqmh}(a). For $H_{c2}\leq H\leq H_{s}$ and $J_{1}:J_{2}:J_{3}=1:2:2$%
, where $H_{s}$ is the saturated magnetic field, the magnetization curve has
the form of

\begin{eqnarray}
m(H) &=&m_{p}+(H-H_{c_{2}})\{k_{c}+(H_{s}-H)[\frac{\beta _{1}}{%
(H-H_{c_{2}})^{1/3}}  \notag \\
&-&\frac{\beta _{2}}{(H_{s}-H)^{1/3}}]\},  \label{eqmh-2}
\end{eqnarray}%
where $k_{c}=(m_{s}-m_{p})/(H_{s}-H_{c_{2}})$ with $m_{s}$ the saturation
magnetization, and $\beta _{1}$, $\beta _{2}$ the parameters. One may see
that when $H=H_{c_{2}}$, $m=m_{p}$; $H=H_{s}$, $m=m_{s}$. A nice fitting to
the DMRG result gives the parameters $H_{c_{2}}/J_{1}=3.15$, $m_{s}=1/2$, $%
H_{s}/J_{1}=4.55$, $\beta _{1}=0.143$, $\beta _{2}=0.178$ and $k_{c}=0.238$,
as shown in Fig. \ref{aaaeqmh}(b). It should be remarked that from the
phenomenological Eqs.(\ref{eqmh-1}) and (\ref{eqmh-2}), we find that, away
from the plateau region, the magnetic field dependence of the magnetization
of this model differs from those of Haldane-type spin chains where $m(H)\sim
(H-H_{c1})^{1/2}$.

\begin{figure}[tbp]
\includegraphics[width = 8.5cm]{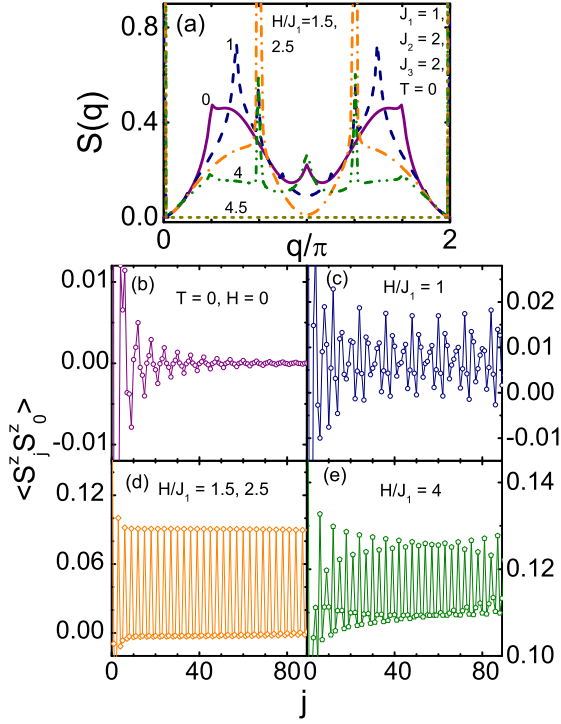}
\caption{ (Color online) For a spin-$1/2$ frustrated Heisenberg diamond
chain with fixed couplings $J_{1}$ : $J_{2}$ : $J_{3}$ = 1 : 2 : 2, (a) the
static structure factor $S(q)$ in the ground states under different external
fields; and the spatial dependence of the spin correlation function $\langle
S^{z}_{j}S^{z}_{0} \rangle$ in the ground states under external field (b) $%
H/J_{1}$ = 0, (c) 1, (d) 1.5 and 2.5, and (e) 4. }
\label{aaasq}
\end{figure}

To explore further the magnetic properties of the frustrated spin-$1/2$
Heisenberg diamond chain in the ground states with the couplings $%
J_{1}:J_{2}:J_{3}=1:2:2$ at different external fields, let us look at the
static structure factor $S(q)$ which is defined as 
\begin{equation}
S(q)=\sum\limits_{j}e^{iqj}\langle S_{j}^{z}S_{0}^{z}\rangle ,  \label{Sq}
\end{equation}%
where $q$ is the wave vector, and $\langle S_{j}^{z}S_{0}^{z}\rangle $ is
the spin correlation function in the ground state. As demonstrated in Fig. %
\ref{aaasq}(a), in the absence of the external field, $S(q)$ shows three
peaks: two at $q=\pi /3$, $5\pi /3$, and one at $q=\pi $, which is quite
different from that of the spin $S=1/2$ Heisenberg AF chain, where $S(q)$
only diverges at $q=\pi $. As indicated by Eq. (\ref{Sq}), the peaks of $%
S(q) $ reflect the periods of the spin correlation function $\langle
S_{j}^{z}S_{0}^{z}\rangle $, i.e., the peaks at $q=\pi /3$ ($5\pi /3$) and $%
\pi $ reflect the periods of $6$ and $2$ for $\langle
S_{j}^{z}S_{0}^{z}\rangle $, respectively. As shown in Fig. \ref{aaasq}(b),
in the absence of the external field, $\langle S_{j}^{z}S_{0}^{z}\rangle $
changes sign every three sites, which corresponds really to the periods of $%
6 $ and $2$. With increasing the magnetic field, the small peak of $S(q)$ at 
$q=\pi $ becomes a round valley while the peak at $q=\pi /3$ ($5\pi /3$)
continuously shifts towards $q=2\pi /3$ ($4\pi /3$) with the height
enhanced, indicating the corruption of the periods of $6$ and $2$ but the
emergence of the new period $\in $ $(3,6)$ for $\langle
S_{j}^{z}S_{0}^{z}\rangle $, as shown in Fig. \ref{aaasq}(c). At the field $%
H/J_{1}=1.5$, two peaks shift to $q=2\pi /3$ and $4\pi /3$ respectively, and
merge into the peaks already existing there, showing the existence of the
period 3 for $\langle S_{j}^{z}S_{0}^{z}\rangle $, as clearly displayed in
Fig. \ref{aaasq}(d). The valley and peaks of $S(q)$ keep intact in the
plateau state at $m=1/6$. When the plateau state is destroyed at the field $%
H/J_{1}=4$, the peaks at $q=2\pi /3$ and $4\pi /3$ are depressed
dramatically while the peaks at $q=\pi /3$, $5\pi /3$ and $\pi $ appear
again with very small heights, revealing the absence of the period $3$ and
the slight presence of the period $2$ and $6$, as shown in Fig. \ref{aaasq}%
(e). At the field $H/J_{1}=4.8$, all peaks disappear and become zero, except
for the peak at $q=0$, which is the saturated state. Therefore, the static
structure factor $S(q)$ shows different characteristics in different
magnetic fields\cite{note1}. On the other hand, it is known that $S(q)$ also
reflects the low-lying excitations of the system. It is thus reasonable to
expect that the low-lying excitations of the frustrated diamond chain will
behave differently in different magnetic fields.

\begin{figure}[tbp]
\includegraphics[width = 8.5cm]{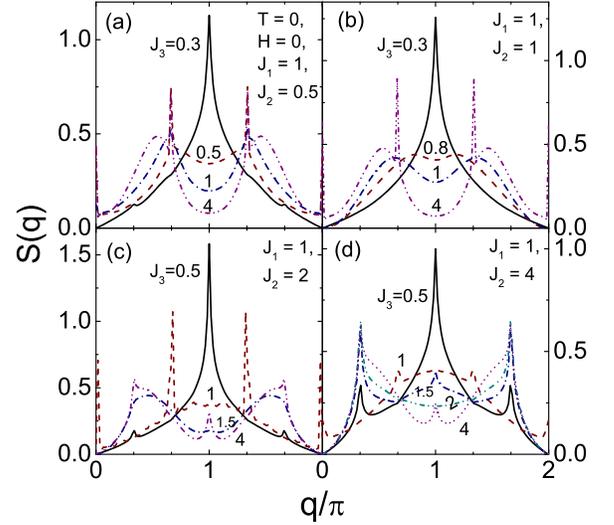}
\caption{ (Color online) The zero-field static structure factor $S(q)$ in
the ground states for the spin-$1/2$ frustrated Heisenberg diamond chains
with length $L=120$, $J_{1}$ = 1, $J_{3}>0$ and $J_{2}$ taken as (a) 0.5,
(b) 1, (c) 2 and (d) 4.}
\label{aaah0}
\end{figure}

To investigate the zero-field static structure factor $S(q)$ in the ground
state for the frustrated spin-$1/2$ diamond chains with various AF
couplings, the four cases with $J_{1}=1$, $J_{3}>0$, and $J_{2}=0.5$, $1$, $%
2 $, and $4$ are shown in Figs. \ref{aaah0}(a)-\ref{aaah0}(d), respectively.
For $J_{2}=0.5$, as shown in Fig. \ref{aaah0}(a), $S(q)$ displays a sharp
peak at $q=\pi $ when $J_{3}<0.5$, and three peaks at $q=0$, $2\pi /3$ and $%
4\pi /3$ when $J_{3}>0.5$. It is shown from the ground state phase diagram 
\cite{aaa2} that the system is in the spin fluid (SF) phase when $J_{1}=1$, $%
J_{2}=0.5$, $J_{3}<0.5$, and enters into the ferrimagnetic (FRI) phase when $%
J_{1}=1$, $J_{2}=0.5$, $J_{3}>0.5$. For $J_{2}=1$, as indicated in Fig. \ref%
{aaah0}(b), the incommensurate peaks exist, such as the case of $J_{3}=0.8$,
where the system is in the dimerized (D) phase \cite{aaa2}. For $J_{2}=2$,
as manifested in Fig. \ref{aaah0}(c), $S(q)$ has a sharp peak at $q=\pi $
and two ignorable peaks at $\pi /3$ and $5\pi /3$ when $J_{3}$ $<1$; three
sharp peaks at $q=0$, $2\pi /3$ and $4\pi /3$ at $J_{3}=1$; a round valley ($%
J_{3}=1.5$) or a small peak ($J_{3}=4$) at $q=\pi $ and two mediate peaks at 
$\pi /3$ and $5\pi /3$ when $J_{3}>1$. The system with $J_{1}=1$ and $%
J_{2}=2 $ is in the D phase when $1<J_{3}<2.8$, and in the SF phase when $%
J_{3}<1$ or $J_{3}>2.8$ \cite{aaa2}. For $J_{2}=4$ revealed in Fig. \ref%
{aaah0}(d), the situations are similar to that of Fig. \ref{aaah0}(c), but
here only the SF phase exists for the system with $J_{1}=1$, $J_{2}=4$ and $%
J_{3}>0$ \cite{aaa2}. It turns out that even in the same phase, such as the
SF phase, the zero-field static structure factor $S(q)$ could display
different characteristics for different AF couplings. In fact, we note that
the exotic peak of $S(q)$ has been experimentally observed in the
diamond-typed compound Sr$_{3}$Cu$_{3}$(PO$_{4}$)$_{4}$ \cite{SrCu}.

If the spin correlation function for the spin-S chain can be expressed as $%
\langle S_{j}^{z}S_{0}^{z}\rangle =\alpha (-1)^{j}e^{-j\beta }$, where $%
\alpha $ and $\beta $ are two parameters, its static structure factor will
take the form of 
\begin{equation}
S(q)=\frac{S(S+1)}{3}-\frac{\alpha (\cos {q}+e^{-\beta })}{\cos {q}+\cosh {%
\beta }},  \label{seq}
\end{equation}%
which can recover exactly the $S(q)$ of the spin-S AKLT chain $S(q)=\frac{S+1%
}{3}\frac{1-\cos {q}}{1+\cos {q}+2/S(S+2)}$ \cite{SQAKLT} with $\alpha
=(S+1)^{2}/3$ and $\beta =\ln (1+2/S)$. Eq.(\ref{seq}) has a peak at
wavevector $q=\pi $. By noting that the zero-field static structure factor $%
S(q)$ for the frustrated diamond chains displays peaks at wavevectors $%
q=a\pi /3$ $(a=0,1,2,3,4,5)$ for different AF couplings, the spin
correlation function $\langle S_{j}^{z}S_{0}^{z}\rangle $ could be
reasonably divided into six modes $\langle S_{6m+l}^{z}S_{0}^{z}\rangle
=c_{l}+\alpha _{l}e^{-(6m+l)\beta }$ or $\langle
S_{6m+l}^{z}S_{0}^{z}\rangle =\alpha _{l}(6m+l)^{-\beta }$ with $j=6m+l$ and 
$l=1,2,...,6$, whose contributions to the static structure factor should be
considered separately\cite{note2}. Thus, the static structure factor for the
present systems could be mimicked by a superposition of six modes, which
leads to 
\begin{eqnarray}
S(q) &=&\sum\limits_{l=1}^{6}[\alpha _{l}\frac{e^{(6-l)\beta }\cos
(lq)-e^{-l\beta }\cos [(6-l)q]}{\cosh (6\beta )-\cos (6q)}  \notag \\
&&+c_{l}\frac{\cos (lq)-\cos [(6-l)q]}{1-\cos (6q)}]+\frac{1}{4},
\label{sixeq}
\end{eqnarray}%
or 
\begin{equation}
S(q)=\sum\limits_{l=1}^{6}\alpha _{l}\sum\limits_{m=0}^{\infty
}2(6m+l)^{-\beta }\cos [(6m+1)q]+\frac{1}{4},  \label{sixeqpow}
\end{equation}%
respectively, depending on which phase the system falls into, where $\alpha
_{l}$, $c_{l}$ and $\beta $ are couplings-dependent parameters.

\begin{figure}[tbp]
\includegraphics[width = 8.5cm]{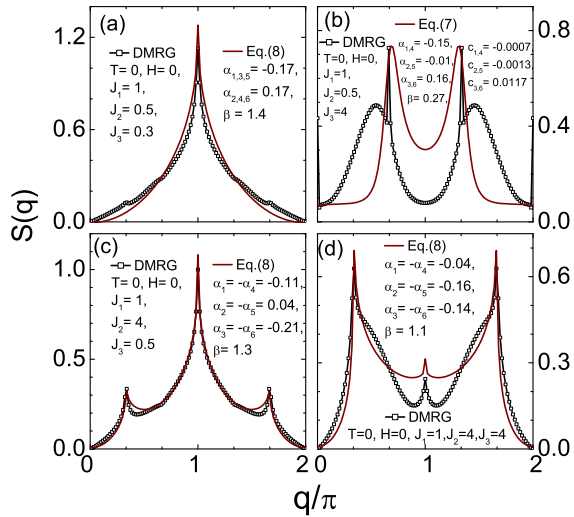}
\caption{ (Color online) The DMRG results of the zero-field static structure
factor as a function of wavevector for the spin-$1/2$ frustrated Heisenberg
diamond chains are fitted: (a) $J_{1}=1$, $J_{2}=0.5$, $J_{3}=0.3$ by Eq. (%
\protect\ref{sixeqpow});   (b) $J_{1}=1$, $J_{2}=0.5$, $J_{3}=4$  by Eq. (%
\protect\ref{sixeq}); (c) $J_{1}=1$, $J_{2}=4$, $J_{3}=0.5$, and (d) $J_{1}=1
$, $J_{2}=4$, $J_{3}=4$ by Eq. (\protect\ref{sixeqpow}). }
\label{aaaeq}
\end{figure}

As presented in Fig. \ref{aaaeq}, the DMRG results of the static structure
factor as a function of wavevector are fitted by Eqs. (\ref{sixeq}) and (\ref%
{sixeqpow}) for the spin-$1/2$ frustrated Heisenberg diamond chains with
various AF couplings in zero magnetic field. It can be found that the
characteristic peaks can be well fitted by Eqs. (\ref{sixeq}) and (\ref%
{sixeqpow}), with only a slightly quantitative deviation, showing that the
main features of the static structure factor for the present systems can be
reproduced by a linear superposition of six modes. The fitting results give
six different modes in general, as shown in Fig. \ref{aaasq}(b).

\begin{figure}[tbp]
\includegraphics[width = 8.5cm]{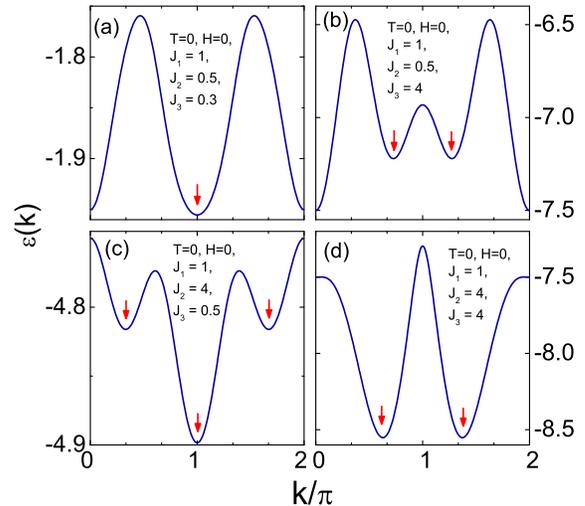}
\caption{ (Color online) The zero-field low-lying fermionic excitation as a
function of wavevector for the spin-$1/2$ frustrated Heisenberg diamond
chain with (a) $J_{1}=1$, $J_{2}=0.5$, $J_{3}=0.3$, (b) $J_{1}=1$, $%
J_{2}=0.5 $, $J_{3}=4$, (c) $J_{1}=1$, $J_{2}=4$, $J_{3}=0.5$, and (d) $%
J_{1}=1$, $J_{2}=4$, $J_{3}=4$. The arrows indicate the locations of
minimums of $\protect\varepsilon (k)$.}
\label{aaaexcita}
\end{figure}

To further understand the above-mentioned behaviors of the zero-field static
structure factor, $S(q)$, we have applied the Jordan-Wigner (JW)
transformation to study the low-lying excitations of the spin-$1/2$
frustrated Heisenberg diamond chain with various AF couplings (see Appendix
A for derivations). It can be seen that the zero-field low-lying fermionic
excitation $\varepsilon (k)$ behaves differently for different AF couplings,
as shown in Figs. \ref{aaaexcita}(a)-(d). Obviously, these low-lying
excitations are responsible for the DMRG calculated behaviors of S(q), where
the positions of minimums of $\varepsilon (k)$ for different AF couplings,
as indicated by arrows in Figs. \ref{aaaexcita}(a)-(c), are exactly
consistent with the locations of the peaks of zero-field static structure
factor $S(q)$ shown in Figs. \ref{aaaeq}(a)-(c), respectively, although
there is a somewhat deviation for Fig. \ref{aaaexcita}(d) and Fig. \ref%
{aaaeq}(d). It also shows that the six modes suggested by Eqs. (\ref{sixeq})
and (\ref{sixeqpow}) is closely related to the low-lying excitations of the
system.

\subsection{Magnetization, Susceptibility and Specific Heat}

\begin{figure}[tbp]
\includegraphics[width = 8.5cm]{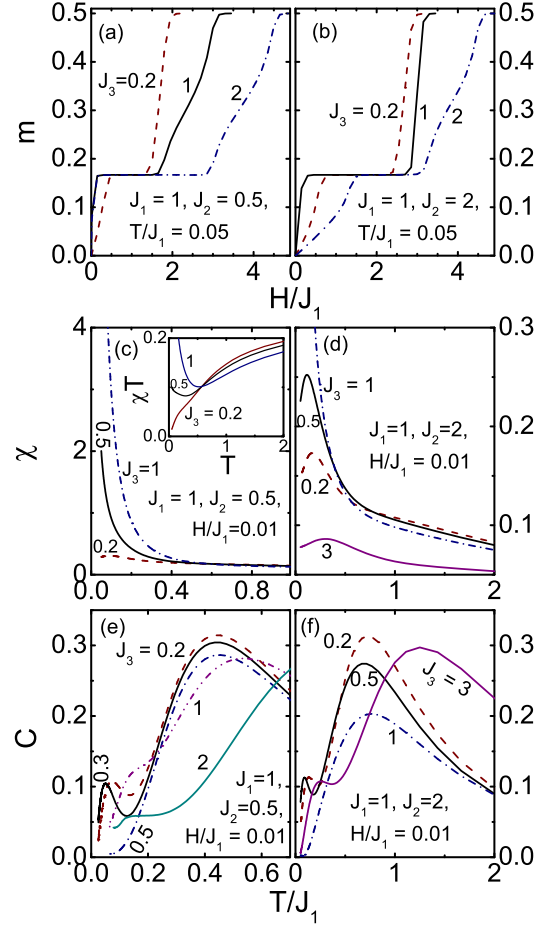}
\caption{ (Color online) For the spin-$1/2$ frustrated Heisenberg diamond
chains with $J_{1}=1$ and $J_{3}>0$, the magnetization process $m(H)$ at
temperature $T/J_{1}=0.05$ with (a) $J_{2}=0.5$ and (b) $J_{2}=2$; the
susceptibility $\protect\chi (T)$ at field $H/J_{1}=0.01$ with (c) $%
J_{2}=0.5 $ and (d) $J_{2}=2$; the specific heat $C(T)$ at field $%
H/J_{1}=0.01$ with (e) $J_{2}=0.5$ and (f) $J_{2}=2$. }
\label{aaatmrg1}
\end{figure}

The magnetization process for the spin-$1/2$ frustrated diamond chain with $%
J_{1}=1$, $J_{3}>0$, and $J_{2}=0.5$ and $2$ is shown in Fig. \ref{aaatmrg1}%
(a) and \ref{aaatmrg1}(b), respectively, where temperature is fixed as $%
T/J_{1}=0.05$. It is found that the magnetization exhibits different
behaviors for different AF couplings: a plateau at $m=1/6$ is observed, in
agreement with the ground state phase diagram \cite{aaa4,aaa5}; for $%
J_{2}=0.5$, as shown in Fig. \ref{aaatmrg1}(a), the larger $J_{3}$ is, the
larger the width of the plateau at $m=1/6$ becomes; for $J_{2}=2$, as
presented in Fig. \ref{aaatmrg1}(b), the width of the plateau at $m=1/6$
becomes larger with increasing $J_{3}<1$, and then turns smaller with
increasing $J_{3}>1$; for $J_{3}<1$, the larger $J_{2}$, the larger the
width of the plateau at $m=1/6$; for $J_{3}=2$, the larger $J_{2}$, the
smaller the width of the plateau at $m=1/6$. The saturated field is
obviously promoted with increasing AF $J_{3}$ and $J_{2}$.

Figures \ref{aaatmrg1}(c) and \ref{aaatmrg1}(d) give the susceptibility $%
\chi $ as a function of temperature $T$ for the spin-$1/2$ frustrated
diamond chain with $J_{1}=1$, $J_{3}>0$ and $J_{2}$ $=0.5$ and $2$,
respectively, while the external field is taken as $H/J_{1}=0.01$. For $%
J_{2}=0.5$, as shown in Fig. \ref{aaatmrg1}(c), the low temperature part of $%
\chi (T)$ keeps finite when $J_{3}<0.5$, and becomes divergent when $%
J_{3}>0.5$. As clearly manifested in the inset of Fig. \ref{aaatmrg1}(c), $%
J_{3}=0.5$ is the critical value, which is consistent with the behaviors of
static structure factor $S(q)$ in Fig. \ref{aaah0}(a). For $J_{2}=2$, as
shown in Fig. \ref{aaatmrg1}(d), an unobvious double-peak structure at low
temperature is observed at small and large $J_{3}$ such as $0.2$ and $3$.
The temperature dependence of the specific heat $C$ with $J_{1}=1$, $J_{3}>0$
and $J_{2}=0.5$ and $2$ is shown in Fig. \ref{aaatmrg1}(e) and \ref{aaatmrg1}%
(f), respectively, while the external field is fixed as $H/J_{1}=0.01$. For $%
J_{2}=0.5$, as given in Fig. \ref{aaatmrg1}(e), a double-peak structure of $%
C(T)$ is observed at low temperature for small and large $J_{3}$ such as $%
0.3 $ and $1$. The case with $J_{2}=2$ shown in Fig. \ref{aaatmrg1}(f)
exhibits the similar characteristics. Thus, the thermodynamics of the system
demonstrate different behaviors for different AF couplings. As manifested in
Fig. \ref{aaah0}, the low-lying excitations behave differently for different
AF couplings. The double-peak structure of the susceptibility as well as the
specific heat could be attributed to the excited gaps in the low-lying
excitation spectrum\cite{GuSu2}.

\section{A Frustrated Diamond Chain with Competing Interactions ($%
J_{1},J_{3}<0,J_{2}>0$)}

\subsection{Local Magnetic Moment and Spin Correlation Function}

\begin{figure}[tbp]
\includegraphics[width = 8.5cm]{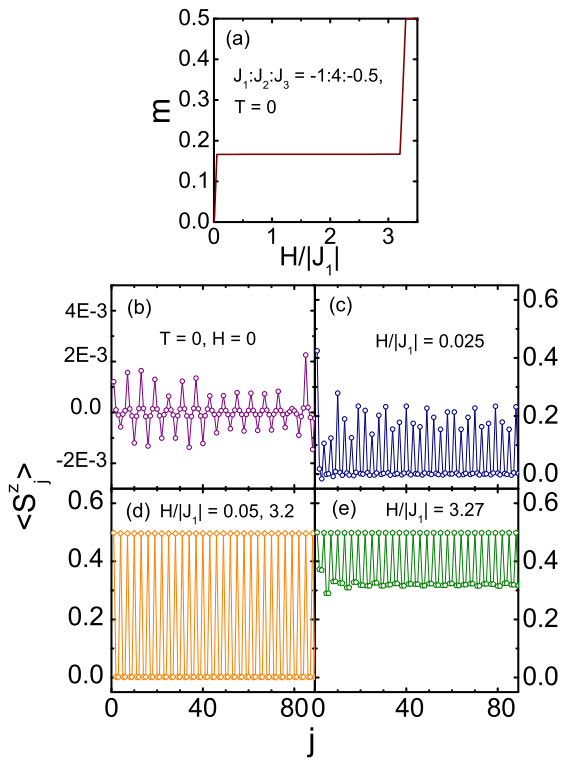}
\caption{ (Color online) For a spin-$1/2$ frustrated Heisenberg diamond
chain with fixed couplings $J_{1}:J_{2}:J_{3}=-1:4:-0.5$, (a) the
magnetization per site $m$ as a function of magnetic field $H$ in the ground
states; and the spatial dependence of the averaged local magnetic moment $%
\langle S_{j}^{z}\rangle $ in the ground states with external field (b) $%
H/|J_{1}|=0$, (c) $0.025$, (d) $0.05$ and $3.2$, and (e) $3.27$. }
\label{fafsj}
\end{figure}

Figure \ref{fafsj}(a) shows the magnetization process of a frustrated spin-$%
1/2$ Heisenberg diamond chain in the ground states with the couplings $%
J_{1}:J_{2}:J_{3}=-1:4:-0.5$. The plateau of magnetization per site $m=1/6$
is clearly obtained. To understand the occurrence of the magnetization
plateau, the spatial dependence of the averaged local magnetic moment $%
\langle S_{j}^{z}\rangle $ in the ground states at different external fields
is calculated. It is seen that in the absence of the magnetic field, as
presented in Fig. \ref{fafsj}(b), the expectation values of $\langle
S_{j}^{z}\rangle $ change sign every three sites within a very small range
of $(-2\times 10^{-4},2\times 10^{-4})$, resulting in the magnetization per
site $m=0$. With increasing the field, the expectation values of $\langle
S_{j}^{z}\rangle $ increase, whose unit of three spins is gradually divided
into a pair and a single, as displayed in Fig. \ref{fafsj}(c). At the field $%
H/|J_{1}|=0.05$, as illustrated in Fig. \ref{fafsj}(d), the behavior of $%
\langle S_{j}^{z}\rangle $ shows a perfect sequence such as $%
\{...,(S_{a},S_{b},S_{b}),...\}$ with $S_{a}=0.496$ and $S_{b}=0.002$,
giving rise to the magnetization per site $m=1/6$. In addition, the sequence
is fixed with increasing the field until $H/|J_{1}|=3.2$, corresponding to
the plateau of $m=1/6$. As the field is enhanced further, the double $S_{b}$
begin to rise, and the sequence becomes a waved series with smaller swing of 
$(S_{a}-S_{b})$ as revealed in Fig. \ref{fafsj}(e), which corresponds to the
plateau state at $m=1/6$ that is destroyed. It is noting that the increase
of $m$ is mainly attributed to the promotion of double $S_{b}$, as $S_{a}$
is already saturated until $S_{a}=S_{b}=0.5$ at the saturated field.

As discussed above, the physical picture of the $m=1/6$ plateau state at $%
J_{1}:J_{2}:J_{3}=-1:4:-0.5$ can be understood by the following approximate
wave function 
\begin{equation*}
\psi _{i}=\frac{1}{\sqrt{2}}(|\uparrow _{3i-2}\uparrow _{3i-1}\downarrow
_{3i}\rangle \pm |\uparrow _{3i-2}\downarrow _{3i-1}\uparrow _{3i}\rangle ).
\end{equation*}%
By use of this wave function, we have $\langle \psi _{i}|S_{3i-2}^{z}|\psi
_{i}\rangle =1/2$, $\langle \psi _{i}|S_{3i-1}^{z}|\psi _{i}\rangle =0$, $%
\langle \psi _{i}|S_{3i}^{z}|\psi _{i}\rangle =0$, leading to a sequence of $%
\{...,(1/2,0,0),...\}$, and $m=(1/2+0+0)/3=1/6.$ This is in agreement with
our DMRG results $\{...,(0.496,0.002,0.002),...\}$.

\begin{figure}[tbp]
\includegraphics[width = 8.5cm]{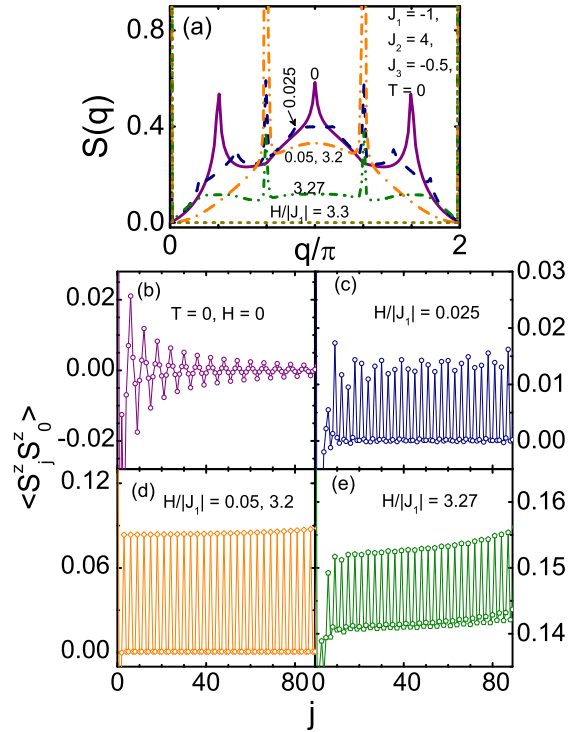}
\caption{ (Color online) For a spin-$1/2$ frustrated Heisenberg diamond
chain with fixed couplings $J_{1}:J_{2}:J_{3}=-1:4:-0.5$, (a) the static
structure factor $S(q)$ in the ground states with different external fields;
and the spatial dependence of the spin correlation function $\langle
S_{j}^{z}S_{0}^{z}\rangle $ in the ground states with external field (b) $%
H/|J_{1}|=0$, (c) $0.025$, (d) $0.05$ and $3.2$, and (e) $3.27$.}
\label{fafsq}
\end{figure}

The static structure factor $S(q)$ of the frustrated spin-$1/2$ Heisenberg
diamond chain with the competing couplings $J_{1}:J_{2}:J_{3}=-1:4:-0.5$ in
the ground states is considered in different external fields. At zero field,
shown in Fig. \ref{fafsq}(a), $S(q)$ has three peaks at $q=\pi /3$, $5\pi /3$
and $\pi $ with mediate heights, which reflects the periods of $6$ and $2$
for $\langle S_{j}^{z}S_{0}^{z}\rangle $, respectively. As shown in Fig. \ref%
{fafsq}(b), in the absence of the external field, $\langle
S_{j}^{z}S_{0}^{z}\rangle $ changes sign every three sites, corresponding to
the periods of $6$ and $2$. With increasing the field, the peak at $q=\pi $
becomes flat with height depressed forwardly, while the peak at $q=\pi /3$ ($%
5\pi /3$) is divided into two peaks shifting oppositely from $q=\pi /3$ ($%
5\pi /3$) with height decreased, indicating the corruption of the periods of 
$6$ and $2$ and the emergence of new periods for $\langle
S_{j}^{z}S_{0}^{z}\rangle $, as shown in Fig. \ref{fafsq}(c). At the field $%
H/|J_{1}|=0.05$, two shifting peaks have respectively reached $q=2\pi /3$
and $4\pi /3$, and are merged with the existing peaks, showing the
occurrence of period $3$ for $\langle S_{j}^{z}S_{0}^{z}\rangle $, as
clearly displayed in Fig. \ref{fafsq}(d). The flat and peaks keep constant
during the plateau state at $m=1/6$. When the plateau state is destroyed at
field $H/J_{1}=3.27$, the peaks at $q=2\pi /3$ and $4\pi /3$ are depressed
sharply, revealing the decay of the period $3$, as shown in Fig. \ref{fafsq}%
(e). At the saturated field of $H/J_{1}=3.3$, all peaks disappear and become
flat with the value zero, which is the saturated state. So, the static
structure factor $S(q)$ shows different characteristics in different
magnetic fields. Similar to the discussions in Fig. \ref{aaasq}, the
low-lying excitations of this frustrated diamond chain would also behave
differently in different magnetic fields.

\begin{figure}[tbp]
\includegraphics[width = 8.5cm]{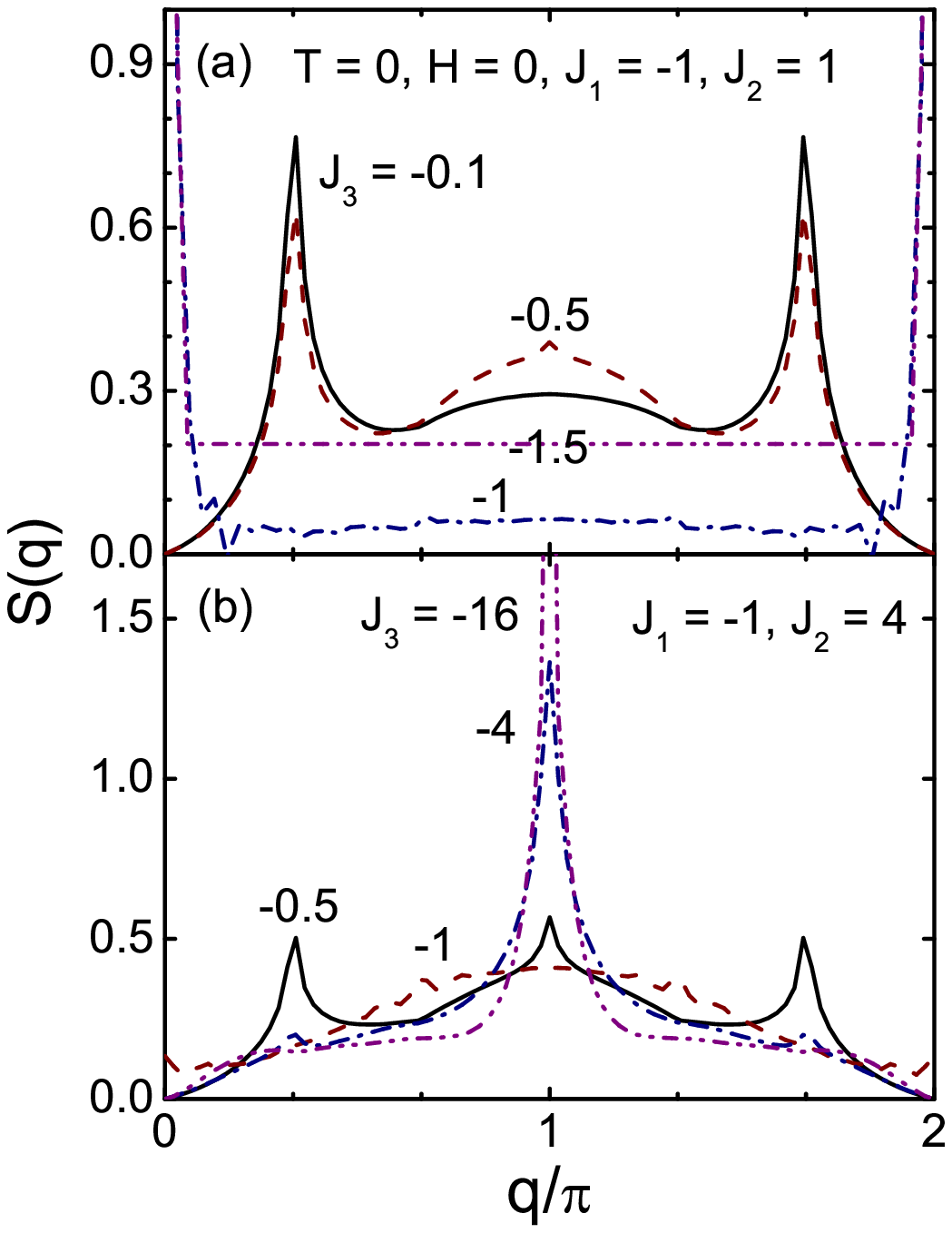}
\caption{ (Color online) The zero-field static structure factor $S(q)$ in
the ground states for the spin-$1/2$ frustrated Heisengberg diamond chains
with length $L=120$, $J_{1}=-1$, $J_{3}<0$ and $J_{2}$ taken as (a) $1$; (b) 
$4$.}
\label{fafh0}
\end{figure}

To further investigate the zero-field static structure factor $S(q)$ in the
ground state for the frustrated spin-$1/2$ diamond chains with various $%
J_{1} $, $J_{3}<0$ and $J_{2}>0$, two cases with $J_{1}=-1$, $J_{3}<0$ and $%
J_{2}=1 $ and $4$ are illustrated in Fig. \ref{fafh0}(a) and \ref{fafh0}(b),
respectively. For $J_{2}=1$, $S(q)$ shows a round peak at $q=\pi $, two
sharp peaks at $\pi /3$ and $5\pi /3$ when $|J_{3}|<1$, and a very sharp
peak at $q=0$ when $|J_{3}|>1$. For $J_{2}=4$, $S(q)$ shows three peaks at $%
q=\pi /3$, $\pi $ and $5\pi /3$ as $|J_{3}|<1$, and a very sharp peak at $%
q=\pi $ and nearly ignorable peaks at $\pi /3$ and $5\pi /3$ as $|J_{3}|>1$.
In general, the zero-field static structure factor $S(q)$ shows different
characteristics with different competing couplings, whose exotic
characteristics could be experimentally observed in the related
diamond-typed compounds.

\begin{figure}[tbp]
\includegraphics[width = 8.5cm]{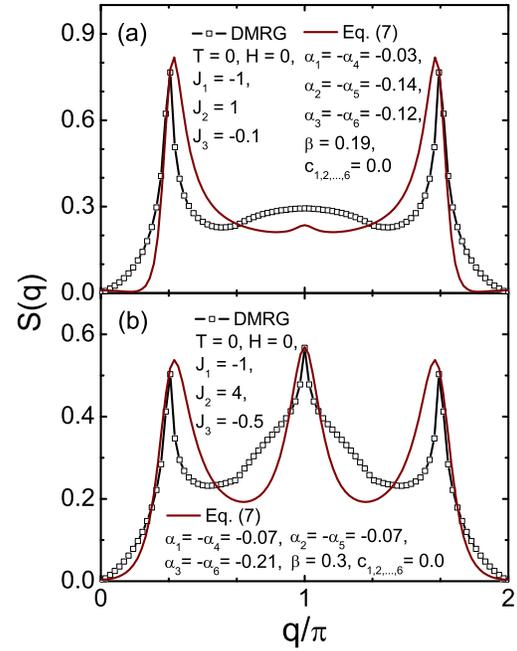}
\caption{ (Color online) The DMRG results of the zero-field static structure
factor as a function of wavevector are fitted by Eq. (\protect\ref{sixeq})
for the spin-$1/2$ frustrated Heisenberg diamond chains with (a) $J_{1}=-1$, 
$J_{2}=1$, $J_{3}=-0.1$, and (b) $J_{1}=-1$, $J_{2}=4$, $J_{3}=-0.5$. }
\label{fafeq}
\end{figure}

As shown in Fig. \ref{fafeq}, the DMRG results of the static structure
factor as a function of wavevector are fitted by Eq.(\ref{sixeq}) for the
spin-$1/2$ frustrated Heisenberg diamond chains with $J_{1}$, $J_{3}<0$ and $%
J_{2}>0$. It can be found that the characteristic behaviors can be nicely
fitted by Eq.(\ref{sixeq}), with only a slightly quantitative deviation,
showing that the main features of the static structure factor for the
present systems can be captured by a superposition of six modes. It is
consistent with the fact that the spin correlation function for the present
systems has six different modes, as manifested in Fig. \ref{fafsq}(b).

\begin{figure}[tbp]
\includegraphics[width = 8.5cm]{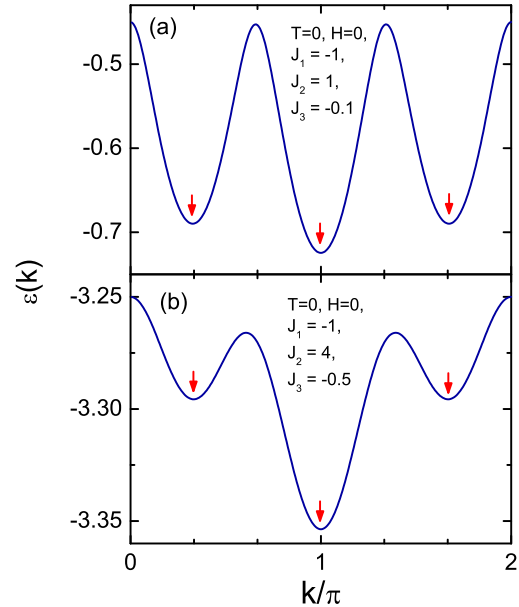}
\caption{ (Color online) The zero-field low-lying fermionic excitation as a
function of wavevector for the spin-$1/2$ frustrated Heisenberg diamond
chain with (a) $J_{1}=-1$, $J_{2}=1$, $J_{3}=-0.1$, and (b) $J_{1}=-1$, $%
J_{2}=4$, $J_{3}=-0.5$. The arrows indicate the locations of minimums of $%
\protect\varepsilon (k)$.}
\label{fafexcita}
\end{figure}

\emph{\ }Similar to the case (a) with all AF couplings in the last section,
the characteristics of zero-field $S(q)$ for the spin-$1/2$ frustrated
Heisenberg diamond chain with $J_{1}$, $J_{3}<0$ and $J_{2}>0$ can be
further undersood in terms of the low-lying excitations of the system (see
Appendix A). By means of the JW transformation, the zero-field low-lying
fermionic excitation $\varepsilon (k)$ of the present case is calculated, as
shown in Figs. \ref{fafexcita}(a)-(b). It is found that the zero-field
low-lying fermionic excitation $\varepsilon (k)$ differs for different
couplings, but the positions of minimums of $\varepsilon (k)$ for different
couplings, as indicated by arrows in Figs. \ref{fafexcita}(a) and (b),
appear to be the same. One may see that these positions coincide exactly
with the locations of peaks of the zero-field static structure factor $S(q)$
manifested in Figs. \ref{fafeq}(a) and (b), respectively, showing that our
fitting equation is qualitatively consistent with the low-lying excitations
of the system.

\subsection{Magnetization, Susceptibility and Specific Heat}

\begin{figure}[tbp]
\includegraphics[width = 8.5cm]{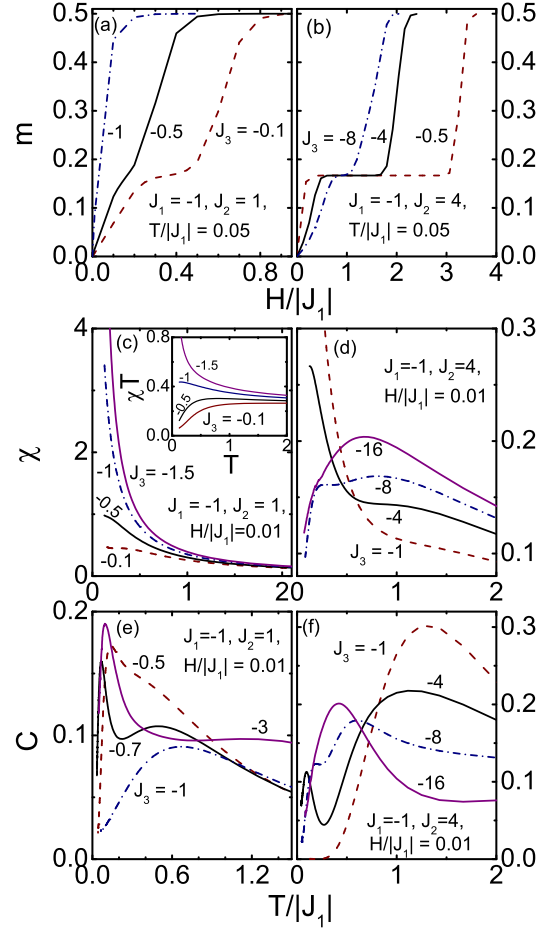}
\caption{ (Color online) For the spin-$1/2$ frustrated Heisenberg diamond
chains with $J_{1}=-1$ and $J_{3}<0$, the magnetization process $m(H)$ at
temperature $T/J_{1}=0.05$ with (a) $J_{2}=1$ and (b) $J_{2}=4$; the
susceptibility $\protect\chi (T)$ at field $H/J_{1}=0.01$ with (c) $J_{2}=1$
and (d) $J_{2}=4$; the specific heat $C(T)$ at field $H/J_{1}=0.01$ with (e) 
$J_{2}=1$ and (f) $J_{2}=4$. }
\label{faftmrg1}
\end{figure}

Figures \ref{faftmrg1}(a) and \ref{faftmrg1}(b) show the magnetization
process for the spin-$1/2$ frustrated diamond chain at a finite temperature $%
T/|J_{1}|=0.05$ with $J_{1}=-1$, $J_{3}<0$, and $J_{2}=1$ and $4$,
respectively. It is shown that the magnetization exhibits different
behaviors for different $J_{1}$, $J_{3}<0$ and $J_{2}>0$. A plateau at $%
m=1/6 $ is observed at small $|J_{3}|$; with a fixed $J_{2}$, the larger $%
|J_{3}|$, the smaller the width of the plateau at $m=1/6$, and the plateau
disappears when $|J_{3}|$ exceeds the critical value; for a fixed $J_{3}$,
the larger $J_{2}$, the wider the width of the plateau at $m=1/6$; the
saturation field is obviously depressed with the increase of $|J_{3}|$ at a
fixed $J_{2}$, and is enhanced with the increase of $J_{2}$ at a fixed $%
J_{3} $.

Figures \ref{faftmrg1}(c) and \ref{faftmrg1}(d) manifest the susceptibility $%
\chi $ as a function of temperature $T$ for the spin-$1/2$ frustrated
diamond chain with $J_{1}=-1$, $J_{3}<0$ and $J_{2}=1$ and $4$,
respectively, where the external field is taken as $H/|J_{1}|=0.01$. For $%
J_{2}=1$, the low temperature part of $\chi (T)$ keeps finite when $%
|J_{3}|<1 $, and becomes divergent when $|J_{3}|>1$. As clearly revealed in
the inset of Fig. \ref{faftmrg1}(c), $J_{3}=-1$ is the critical value, which
is in agreement with the behaviors of static structure factor $S(q)$ shown
in Fig. \ref{fafh0}(a). For $J_{2}=4$, a clear double-peak structure of $%
\chi (T)$ is obtained at $|J_{3}|=8$. The temperature dependence of the
specific heat $C$ with $J_{1}=-1$, $J_{3}<$ $0$ and $J_{2}=1$ and $4$ is
shown in Figs. \ref{faftmrg1}(e) and \ref{faftmrg1}(f), respectively, where
the external field is fixed as $H/|J_{1}|=0.01$. For $J_{2}=1$, a
double-peak structure of $C(T) $ is observed for the case of $|J_{3}|=0.5$.
The case with $J_{2}=4$ shown in Fig. \ref{faftmrg1}(f) exhibits the similar
characteristics. It is also found that, owing to the competitions among $%
J_{1}$, $J_{3}$ and $J_{2}$, the thermodynamics demonstrate rich behaviors
at different couplings. As reflected in Fig. \ref{fafh0}, the low-lying
excitations behave differently with various F interactions $J_{1}$, $J_{3}$
and AF interaction $J_{2}$, while the excitation gaps could induce the
double-peak structure in the susceptibility as well as in the specific heat%
\cite{GuSu2}.

\section{A Diamond Chain without Frustration ($J_{1},J_{2}>0,J_{3}<0$)}

\subsection{Local Magnetic Moment and Spin Correlation Function}

\begin{figure}[tbp]
\includegraphics[width = 8.5cm]{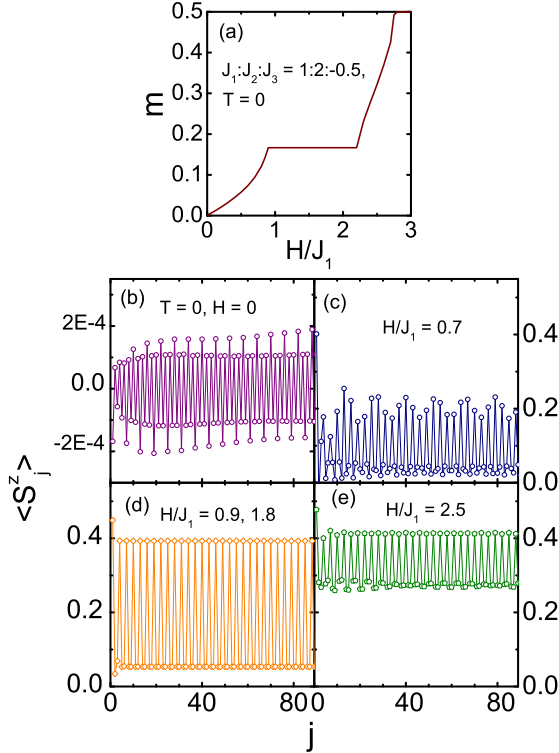}
\caption{ (Color online) For a spin-$1/2$ non-frustrated Heisenberg diamond
chain with fixed couplings $J_{1}:J_{2}:J_{3}=1:2:-0.5$, (a) the
magnetization per site $m$ as a function of magnetic field $H$ in the ground
states; and the spatial dependence of the averaged local magnetic moment $%
\langle S_{j}^{z}\rangle $ in the ground states with external field (b) $%
H/J_{1}=0$, (c) $0.7$, (d) $0.9$ and $1.8$, and (e) $2.5$. }
\label{aafsj}
\end{figure}

Figure \ref{aafsj}(a) shows the magnetization process of a non-frustrated
spin-$1/2$ Heisenberg diamond chain in the ground states with the couplings
satisfying $J_{1}:J_{2}:J_{3}=1:2:-0.5$. The plateau of magnetization per
site $m=1/6$ is observed. The appearance of the magnetization plateau can be
understood from the spatial dependence of the averaged local magnetic moment 
$\langle S_{j}^{z}\rangle $ in the ground states under different external
fields. In absence of the external field, the expectation values of $\langle
S_{j}^{z}\rangle $ change sign every one site with a waved swing within a
very small range of $(-2\times 10^{-4},2\times 10^{-4})$, giving rise to the
magnetization per site $m=0$. Under a finite field, every three successive
spins have gradually cooperated into a pair and a single, as shown in Fig. %
\ref{aafsj}(c). At $H/J_{1}=0.9$, as given in Fig. \ref{aafsj}(d), $\langle
S_{j}^{z}\rangle $ shows a perfect sequence such as $%
\{...,(S_{a},S_{b},S_{b}),...\}$ with $S_{a}=0.393$ and $S_{b}=0.053$,
resulting in the magnetization per site $m=1/6$. Moreover, the sequence is
fixed with the field increased until $H/J_{1}=1.8$, corresponding to the
plateau state of $m=1/6$. As the field is increased further, double $S_{b}$
begin to increase, and the sequence becomes a waved succession with a
smaller swing of $(S_{a}-S_{b})$, as manifested in Fig. \ref{aafsj}(e),
which corresponds to the fact that the plateau state at $m=1/6$ is
destroyed. It is observed that the increase of $m$ at first is mainly
attributed to the speedy boost of double $S_{b}$, and later, $S_{a}$ starts
to increase weakly until $S_{a}=S_{b}=0.5$ at the saturated field.

For this non-frustrated case with couplings $J_{1}:J_{2}:J_{3}=1:2:-0.5$,
the obtained perfect sequence of $\{...,(0.393,0.053,0.053),...\}$ for the $%
m=1/6$ plateau state could be understood by the following approximate
trimerized wave function 
\begin{eqnarray}
\psi _{i} &=&\frac{1}{3}(2|\uparrow _{3i-2}\uparrow _{3i-1}\downarrow
_{3i}\rangle \pm 2|\uparrow _{3i-2}\downarrow _{3i-1}\uparrow _{3i}\rangle 
\notag \\
&\pm &|\downarrow _{3i-2}\uparrow _{3i-1}\uparrow _{3i}\rangle ).
\label{aafwave}
\end{eqnarray}
According to this function, we find $\langle \psi _{i}|S_{3i-2}^{z}|\psi
_{i}\rangle =7/18$, $\langle \psi _{i}|S_{3i-1}^{z}|\psi _{i}\rangle =1/18$, 
$\langle \psi _{i}|S_{3i}^{z}|\psi _{i}\rangle =1/18$,\emph{\ }giving rise
to a sequence of $\{...,(7/18,1/18,1/18),...\}$. It turns out that $%
m=(7/18+1/18+1/18)/3=1/6$. This observation implies that the ground state of
the plateau state can also be described by the trimerized states.

\begin{figure}[tbp]
\includegraphics[width = 8.5cm]{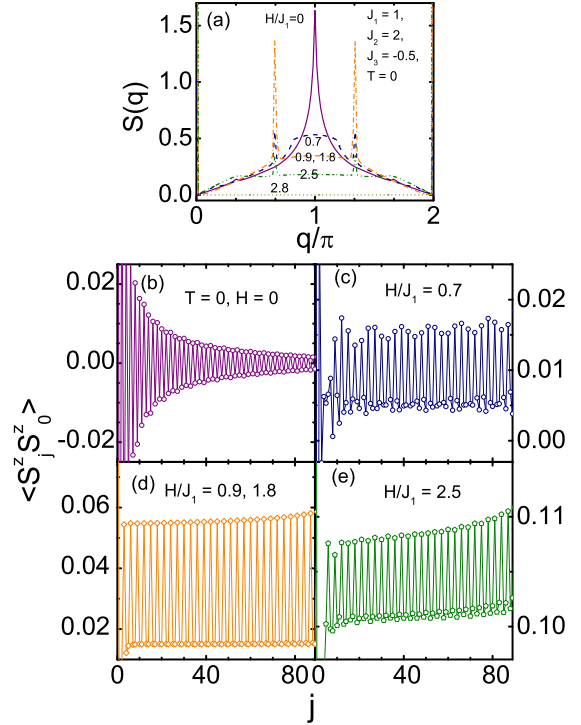}
\caption{ (Color online) For a spin-$1/2$ non-frustrated Heisenberg diamond
chain with fixed couplings $J_{1}:J_{2}:J_{3}=1:2:-0.5$, (a) the static
structure factor $S(q)$ in the ground states with different external fields;
and the spatial dependence of the spin correlation function $\langle
S_{j}^{z}S_{0}^{z}\rangle $ in the ground states with external field (b) $%
H/J_{1}=0$, (c) $0.7$, (d) $0.9$ and $1.8$, and (e) $2.5$.}
\label{aafsq}
\end{figure}

The static structure factor $S(q)$ of the non-frustrated spin-$1/2$
Heisenberg diamond chain in the ground states with the couplings $%
J_{1}:J_{2}:J_{3}=1:2:-0.5$ is probed in different external fields. As shown
in Fig. \ref{aafsq}(a), in absence of the external field, $S(q)$ shows a
sharp peak at $q=\pi $, similar to the behaviors of the $S=1/2$ Heisenberg
AF chain, which reflects the period of $2$ for $\langle
S_{j}^{z}S_{0}^{z}\rangle $. As displayed in Fig. \ref{aafsq}(b), at zero
external field, $\langle S_{j}^{z}S_{0}^{z}\rangle $ changes sign every one
lattice site, corresponding to the period of $2$. When the field is
increased, the peak at $q=\pi $ becomes a flat with the height depressed
greatly, while two new peaks with small heights at $q=2\pi /3$ and $4\pi /3$
appear, indicating the corruption of the period of $2$ and the emergence of
the new period of $3$ for $\langle S_{j}^{z}S_{0}^{z}\rangle $, as
demonstrated in Fig. \ref{aafsq}(c). At the field $H/J_{1}=0.9$, the peaks
at $q=0$, $2\pi /3$ and $4\pi /3$ become sharper. The flat and peaks of $%
S(q) $ keep unchanged during the plateau state at $m$ = $1/6$. When the
plateau state is destroyed at the field $H/J_{1}=2.5$, the peaks at $q=2\pi
/3$ and $4\pi /3$ are suppressed dramatically, revealing the decay of the
period $3$, as shown in Fig. \ref{aafsq}(e). At the field $H/J_{1}=2.8$, all
peaks disappear, and become a flat with the value zero, except for the peak
at $q$ $=0$, which is the saturated state. It can be stated that the static
structure factor $S(q)$ shows various characteristics in different magnetic
fields. Similar to what discussed in Figs. \ref{aaasq} and \ref{fafsq}, the
low-lying excitations of this non-frustrated diamond chain would also behave
differently in different magnetic fields.

\begin{figure}[tbp]
\includegraphics[width = 8.5cm]{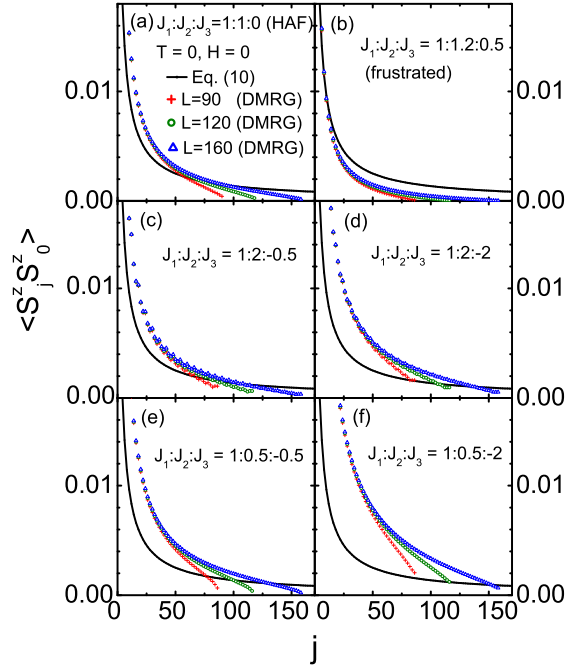}
\caption{ (Color online) The zero-field static correlation function $\langle
S_{j}^{z}S_{0}^{z}\rangle $ versus site $j$ in the ground state for a spin-$%
1/2$ diamond chain with different lengths and various couplings. The
couplings ratio $J_{1}:J_{2}:J_{3}$ is taken as (a) $1:1:0$ (HAF), (b) $%
1:1.2:0.5$ (frustrated), (c) $1:2:-0.5$, (d) $1:2:-2$, (e) $1:0.5:-0.5$, (f) 
$1:0.5:-2$. The length is taken as $L=90$, $120$ and $160$, respectively.}
\label{aafh0}
\end{figure}

The zero-field static structure factor $S(q)$ in the ground state for the
present system displays a peak at $q=\pi $ with different couplings, but
whose static correlation function $\langle S_{j}^{z}S_{0}^{z}\rangle $
varies with the couplings. As illustrated in Fig. \ref{aafh0}, only are the
values of $\langle S_{j}^{z}S_{0}^{z}\rangle $ larger than zero presented
for convenience. In order to gain deep insight into physics, for a
comparison we also include the static correlation function for the $S=1/2$
Heisenberg antiferromagnetic (HAF) chain in Fig. \ref{aafh0}(a), whose
asymptotic behavior has the form of \cite{HAF1, HAF2} 
\begin{equation}
\langle \mathbf{S}_{j}\cdot \mathbf{S}_{0}\rangle \propto (-1)^{j}\frac{1}{%
(2\pi )^{3/2}}\frac{\sqrt{\ln j}}{j}.  \label{equa}
\end{equation}%
Eq. (\ref{equa}) is depicted as solid lines in Figs. \ref{aafh0}. To take
the finite-size effect into account, the length of the diamond chain is
taken as $L=90$, $120$ and $160$, respectively. As revealed in Fig. \ref%
{aafh0}(a), the DMRG result of the $S=1/2$ HAF chain with an infinite length
agrees well with the solid line. Fig. \ref{aafh0}(b) shows that the static
correlation function for the spin-$1/2$ diamond chain with frustrated
couplings $J_{1}:J_{2}:J_{3}=1:1.2:0.5$ decays faster than that of the HAF
chain. Compared with Figs. \ref{aafh0}(c)-(f), it can be found that all the
static correlation functions for the spin-$1/2$ non-frustrated diamond chain
with AF interactions $J_{1}$, $J_{2}$ and F interaction $J_{3}$ fall more
slowly than that of the $S=1/2$ HAF chain; for fixed AF interactions $J_{1}$
and $J_{2}$, the static correlation functions drop more leisurely with
increasing the F interaction $|J_{3}|$; for fixed AF interaction $J_{1}$ and
F interaction $J_{3}$, the static correlation functions decrease more
rapidly with increasing the AF interaction $J_{2}$.

\subsection{Magnetization, Susceptibility and Specific Heat}

\begin{figure}[tbp]
\includegraphics[width = 8.5cm]{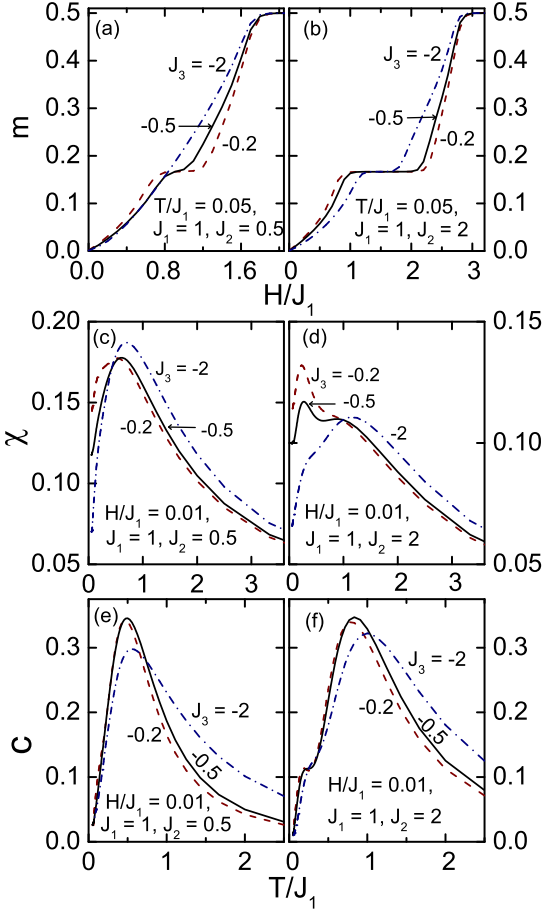}
\caption{ (Color online) For the spin-$1/2$ non-frustrated Heisenberg
diamond chains with $J_{1}=1$ and $J_{3}<0$, the magnetization process $m(H)$
at temperature $T/J_{1}=0.05$ with (a) $J_{2}=0.5$ and (b) $J_{2}=2$; the
susceptibility $\protect\chi (T)$ at field $H/J_{1}=0.01$ with (c) $%
J_{2}=0.5 $ and (d) $J_{2}=2$; the specific heat $C(T)$ at field $%
H/J_{1}=0.01$ with (e) $J_{2}=0.5$ and (f) $J_{2}=2$. }
\label{aaftmrg1}
\end{figure}

Figures \ref{aaftmrg1}(a) and \ref{aaftmrg1}(b) show the magnetization
process for the spin-$1/2$ non-frustrated diamond chain at a finite
temperature $T/J_{1}=0.05$ with $J_{1}=1$, $J_{3}<0$, and $J_{2}=0.5$ and $2$%
, respectively. It is found that the magnetization behaves differently with
different AF interactions $J_{1}$, $J_{2}$ and F interaction $J_{3}$. A
plateau at $m=1/6$ is obtained at small $|J_{3}|$; for fixed $J_{1}$ and $%
J_{2}$, the larger $|J_{3}|$, the narrower the width of the plateau at $%
m=1/6 $, and after $|J_{3}|$ exceeds a critical value, the plateau at $m=1/6$
is eventually smeared out; for fixed $J_{1}$ and $J_{3}$, the larger $J_{2}$%
, the wider the width of the plateau at $m=1/6$; the saturated field is
obviously unchanged with changing the F interaction $J_{3}$. The
coupling-dependence of the spin-$1/2$ non-frustrated diamond chain with AF
interactions $J_{1}$, $J_{2}$ and F interaction $J_{3}$ is similar to that
of trimerized F-F-AF chains\cite{GuSu1}.

Figures \ref{aaftmrg1}(c) and \ref{aaftmrg1}(d) present the susceptibility $%
\chi $ as a function of temperature $T$ for the spin-$1/2$ frustrated
diamond chain with $J_{1}=1$, $J_{3}<0$ and $J_{2}=0.5$ and $2$,
respectively, where the external field is taken as $H/J_{1}=0.01$. A
double-peak structure of $\chi (T)$ is observed at small F interaction $%
|J_{3}|$ and disappears at large $|J_{3}|$. The temperature dependence of
the specific heat $C$ with $J_{1}$ $=1$, $J_{3}<0$ and $J_{2}=0.5$ and $2$
is shown in Figs. \ref{aaftmrg1}(e) and \ref{aaftmrg1}(f), respectively,
where $H/J_{1}=0.01$. It is seen that, when $J_{2}$ is small, $C(T)$
exhibits only a single peak; when $J_{2}$ is large, a double-peak structure
of $C(T)$ is observed. In the latter case, the double-peak structure is more
obvious for small F interaction $|J_{3}|$, and tend to disappear at large $%
|J_{3}|$. Therefore, the thermodynamics demonstrate various behaviors with
different AF interactions $J_{1}$, $J_{2}$ and F interaction $J_{3}$.

\subsection{Effect of Anisotropy of Bond Interactions}

\begin{figure}[tbp]
\includegraphics[width = 8.5cm]{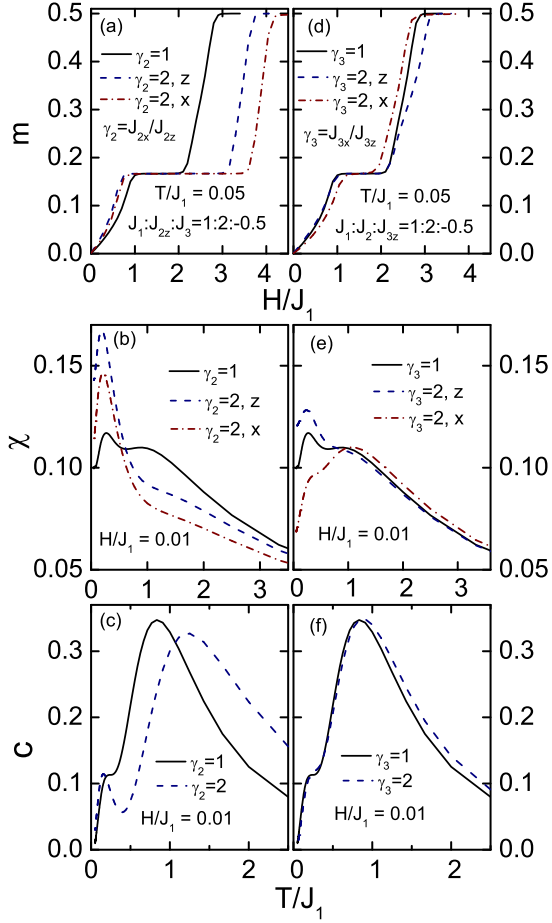}
\caption{ (Color online) For the spin-$1/2$ non-frustrated diamond chain
with the couplings satisfying $J_{1}:J_{2z}:J_{3}=1:2:-0.5$ for various
anisotropy $\protect\gamma _{2}=J_{2x}/J_{2z}\geq 1$: (a) the magnetization
process $m(H)$ at temperature $T/J_{1}=0.05$; (b) the susceptibility $%
\protect\chi (T$ at field $H/J_{1}=0.01$; and (c) the specific heat $C(T)$
at field $H/J_{1}=0.01$. For the spin-$1/2$ non-frustrated diamond chain
with couplings $J_{1}:J_{2}:J_{3z}=1:2:-0.5$ for various anisotropy $\protect%
\gamma _{3}=$ $J_{3x}/J_{3z}\geq 1$: (d) the magnetization process $m(H)$ at
temperature $T/J_{1}=0.05$; (e) the susceptibility $\protect\chi (T)$ at
field $H/J_{1}=0.01$; and (f) the specific heat $C(T)$ at field $%
H/J_{1}=0.01 $.}
\label{fig-tmrg23}
\end{figure}

Some magnetic materials show different behaviors under longitudinal and
transverse magnetic fields, showing that the anisotropy plays an important
role in the physical properties of the system. First, let us investigate the
XXZ anisotropy of the AF interaction $J_{2}$ on the properties of the spin-$%
1/2$ non-frustrated diamond chain with the couplings $%
J_{1}:J_{2z}:J_{3}=1:2:-0.5$ for various anisotropy parameter defined by $%
\gamma _{2}=J_{2x}/J_{2z}=J_{2y}/J_{2z}$, where the $z$ axis is presumed to
be perpendicular to the chain direction. For $\gamma _{2}\geq 1$, the
magnetization $m(H)$, susceptibility $\chi (T)$ and specific heat $C(T)$ are
presented in Figs. \ref{fig-tmrg23}(a), (b) and (c), respectively. With
increasing $\gamma _{2}$, it is found that when the magnetic field $H$ is
along the $z$ direction, the width of the magnetization plateau at $m=1/6$
as well as the saturation field are enlarged, while those are more increased
for $H$ along the $x$ direction than along the $z$ direction; the peak of
the susceptibility $\chi (T)$ for $H$ along the $z$ direction at lower
temperature side is promoted, and the second round peak at high temperature
side is depressed with a little shift, while $\chi (T)$ for $H$ along the $x$
direction shows the similar varying trend; the peak of the specific heat $%
C(T)$ for $H$ along the $z$ direction at lower temperature side leaves
almost unchanged, and the second round peak at high temperature side moves
towards the higher temperature side, while $C(T)$ for $H$ along the $x$
direction coincides with those for $H$ along the $z$ direction. For $%
0<\gamma _{2}<1$, the anisotropy just shows very reverse effect on the
thermodynamic properties in comparison to what we discussed above.

Now let us discuss the effect of the XXZ anisotropy of $J_{3}<0$ on the
magnetic and thermodynamic properties of the spin-$1/2$ non-frustrated
diamond chain with the couplings $J_{1}:J_{2}:J_{3z}=1:2:-0.5$. Recall that
as the $J_{2}$ bond connects two different lattice sites, as shown in Fig. %
\ref{chain}, $J_{1}$ and $J_{3}$ can be different, even their signs. Define
a parameter $\gamma _{3}$ to characterize the anisotropy as $\gamma
_{3}=J_{3x}/J_{3z}=J_{3y}/J_{3z}$, where the $z$ axis is perpendicular to
the chain direction. For $\gamma _{3}\geq 1$, the magnetization $m(H)$,
susceptibility $\chi (T)$ and specific heat $C(T)$ are depicted in Figs. \ref%
{fig-tmrg23}(d), (e) and (f), respectively. With increasing $\gamma _{3}$,
it is seen that the width of the plateau at $m=1/6$ for $H$ along the $z$
direction becomes slightly wider, while it goes smaller for $H$ along the $x$
direction; the saturation field is not changed with $\gamma _{3}$ along both
directions; the peak of $\chi (T)$ for $H$ along the $z$ direction at lower
temperature side is promoted, and the second round peak at higher
temperature side is slightly depressed, while the situations along the $x$
direction are just reverse, namely, the peak at lower temperature side is
depressed, and the second peak at higher temperature side is slightly
promoted; the peak of $C(T)$ for $H$ along the $z$ direction at lower
temperature side leaves almost unchanged, and the second round peak at
higher temperature side moves slightly to the higher temperature side, while 
$C(T)$ along the $x$ direction coincides with that along the $z$ direction.
For $0<\gamma _{3}<1$, the situation just becomes reverse in comparison to
what we discussed above.

\subsection{Comparison to Experimental Results}

Recently, Kikuchi \textit{et al.} \cite{CuCOOH1} have performed a nice
measurement on a spin-$1/2$ diamond-chain compound Cu$_{3}$(CO$_{3}$)$_{2}$%
(OH)$_{2}$, i.e., azurite. They have observed the $1/3$ magnetization
plateau, unambiguously confirming the previous theoretical prediction. The
two broad peaks both in the magnetic susceptibility and the specific heat
are observed. We note that in Ref. \cite{CuCOOH1}, the experimental data at
finite temperatures are fitted by the zero-temperature theoretical results
obtained by the exact diagonalization and DMRG methods, while the result of
the high temperature series expansion fails to fit the low-temperature
behavior of the susceptibility. In accordance with our preceding
discussions, by using the TMRG method, we have attempted to re-analyse the
experimental data presented in Ref. \cite{CuCOOH1} to fit the experiments
for the whole available temperature region.

\begin{figure}[tbp]
\includegraphics[width = 8.5cm]{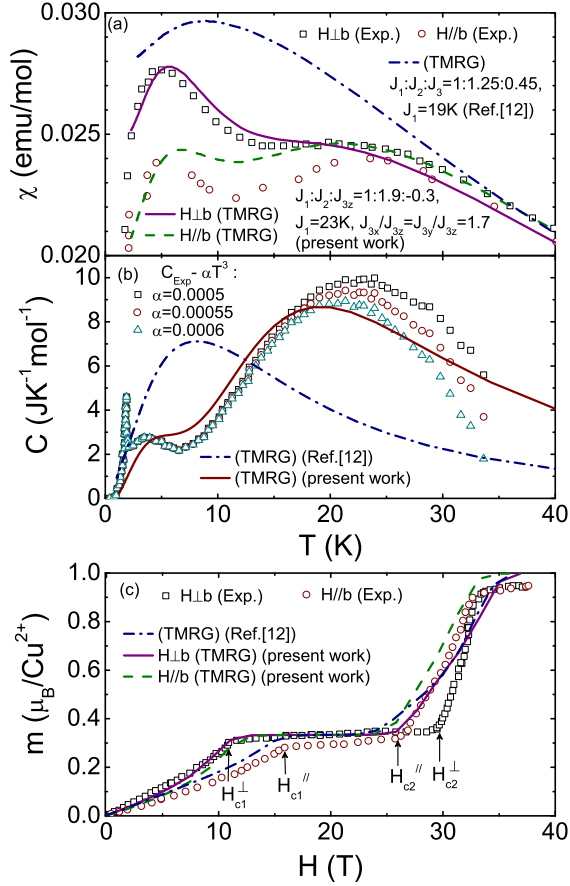}
\caption{ (Color online) A comparison of experimental results for (a) the
magnetic susceptibility, (b) the specific heat and (c) the magnetization
process for the spin-$1/2$ diamond compound Cu$_{3}$(CO$_{3}$)$_{2}$(OH)$%
_{2} $ with the TMRG results. The experimental data are taken from Ref. 
\protect\cite{CuCOOH1}. See the context for details.}
\label{exp}
\end{figure}

Our fitting results for the temperature dependence of the susceptibility $%
\chi $ of the compound Cu$_{3}$(CO$_{3}$)$_{2}$(OH)$_{2}$ are presented in
Fig. \ref{exp}(a). For a comparison, we have also included the TMRG result
calculated by using the parameters given in Ref. \cite{CuCOOH1}. Obviously,
our TMRG results with $J_{1}:J_{2}:J_{3z}=1:1.9:-0.3$ and $%
J_{3x}/J_{3z}=J_{3y}/J_{3z}=1.7$ fit very well the experimental data of $%
\chi $, and the two round peaks at low temperatures are nicely reproduced,
while the result with $J_{1}:J_{2}:J_{3}=1:1.25:0.45$ obtained in Ref. \cite%
{CuCOOH1} cannot fit the low-temperature behavior of $\chi $ \cite{GuSu3}.
On the other hand, the fitting results for the temperature dependence of the
specific heat $C(T)$ of the compound Cu$_{3}$(CO$_{3}$)$_{2}$(OH)$_{2}$ are
shown in Fig. \ref{exp}(b). The lattice contribution, which is included in
the raw experimental data in Ref.\cite{CuCOOH1}, is subtracted according to $%
C(T)=C_{Exp}(T)-\alpha T^{3}$, where $\alpha $ is a parameter. Obviously,
our TMRG result with the same set of parameters $%
J_{1}:J_{2}:J_{3z}=1:1.9:-0.3$ and $J_{3x}/J_{3z}=J_{3y}/J_{3z}=1.7$ fits
also remarkably well the experimental data of $C(T)$, and the two round
peaks at low temperatures are nicely reproduced, while the result with $%
J_{1}:J_{2}:J_{3}=1:1.25:0.45$ given in Ref. \cite{CuCOOH1} cannot fit the
low-temperature behavior of $C(T)$, even qualitatively. In addition, the
sharp peak of $C(T)$ experimentally observed at temperature around $2K$
cannot be reproduced by both sets of the coupling parameters, which might be
a three-dimensional long-range ordering due to interchain interactions. The
fitting results for the magnetization $m(H)$ of the compound Cu$_{3}$(CO$%
_{3} $)$_{2}$(OH)$_{2}$ are shown in Fig. \ref{exp}(c). We would like to
point out that the quantitative fitting by our above parameters to the width
of the plateau is not so good, but the qualitative behavior is quite
consistent with the experiments both in the transverse and longitudinal
magnetic fields, say, $H_{c1}^{\parallel }>H_{c1}^{\perp }$, $%
H_{c2}^{\parallel }<H_{c2}^{\perp }$, and the saturation field is fixed
along both directions, suggesting that our fitting parameters capture the
main characteristics. It is worth pointing out that if the anisotropy ratio
is increased up to $J_{3x}/J_{3z}=J_{3y}/J_{3z}=2.5$ with the same couplings 
$J_{1}:J_{2}:J_{3z}=1:1.9:-0.3$, the width of the $1/3$ plateau for $%
H\parallel b$ will be decreased to about one-half of that for $H\perp b$.

Therefore, our calculations show that (i) the best couplings obtained by
fitting the experimental data of the susceptibility for the azurite could be 
$J_{1}:J_{2}:J_{3z}=1:1.9:-0.3$ with the anisotropic ratio for the
ferromagnetic interaction $J_{3x}/J_{3z}=J_{3y}/J_{3z}=1.7$, where $z\perp b$%
; (ii) the compound may not be a spin frustrated magnet; (iii) the double
peaks of the susceptibility and the specific heat are not caused by the spin
frustration effect, but by the two kind of gapless and gapful excitations
owing to the competition of the AF and F interactions.

One might argue that for this diamond chain compound, from the point of the
lattice distance it is unlikely that $J_{1}$ is AF without XXZ anisotropy
while $J_{3}$ is F with strong XXZ anisotropy. We may offer another
possibility to support our findings, namely, the case of $J_{1}$ and $J_{3}$
with opposite signs is not excluded from the lattice structure of the
compound. A linear relationship exists between the exchange energy and the
metal-ligand-metal bridge angle: the coupling energy, positive
(ferromagnetic) at angles near $90^{o}$, becomes increasingly smaller (more
antiferromagnetic) as the angle increases\cite{Angle}. As the ferromagnetic
coupling $J_{3}$ is determined by fitting the experimental low-temperature
behaviors of $\chi (T)$ and $C(T)$, this fitting coupling parameters should
not be impossible if one considers the angle of $J_{1}$ bridge to keep the
antiferromagnetic coupling while the angle of $J_{3}$ bridge to induce the
ferromagnetic coupling. On the other hand, we note that there is another
compound with Cu ions, Cu$_{2}$(abpt)(SO$_{4}$)$_{2}$(H$_{2}$O)$\cdot $H$%
_{2} $O, whose $g$ factors in XY plane are different from that in $z$
direction \cite{LargeXY}. Besides, someone might argue that the condition $%
J_{2}\gg J_{1},|J_{3}|$ is necessary to explain the double peak behavior of
the diamond chain. In fact, such an argument is not necessarily true, as
manifested in Fig. \ref{aaftmrg1}(c), where the double peaks of $\chi (T)$
at low temperatures can also be produced with the parameters $%
J_{1}:J_{2}:J_{3}=1:0.5:-0.1$. In other words, the double-peak behavior of
the diamond chain may not depend on whether $J_{2}\gg J_{1}$, $J_{3}$ or
not, but may be strongly dependent on the competition of AF and F
interactions, as discussed above.

\section{Summary and discussion}

In this paper, we have numerically studied the magnetic and thermodynamic
properties of spin-$1/2$ Heisenberg diamond chains with three different
cases (a) $J_{1}$, $J_{2}$, $J_{3}>0$ (frustrated), (b) $J_{1}$, $J_{3}<0$, $%
J_{2}>0$ (frustrated), and (c) $J_{1}$, $J_{2}>0$, $J_{3}<0$
(non-frustrated) by means of the DMRG and TMRG methods. In the ground
states, the local magnetic moment, spin correlation function, and static
structure factor are explored. The static structure factor $S(q)$ at zero
field shows peaks at wave vector $q=0$, $\pi /3$, $2\pi /3$, $\pi $, $4\pi /3
$ and $5\pi /3$ for different couplings, in which the peaks at $q=0$, $2\pi
/3$ and $4\pi /3$ in the magnetization plateau state with $m=1/6$ are
observed to be couplings independent. The DMRG results of the zero-field
static structure factor can be nicely fitted by a linear superposition of
six modes, where two fitting equations are proposed. It is seen that the six
modes are closely related to the low-lying excitations of the system. At
finite temperatures, the magnetization, susceptibility and specific heat are
calculated, which show various behaviors for different couplings. The
double-peak structure of the susceptibility and specific heat can be
procured, whose positions and heights are found to be dependent on competing
couplings. It has been shown that the XXZ anisotropy of F and AF couplings
can have remarkable effect on the physical behaviors of the system. In
addition, the experimental susceptibility, specific heat and magnetization
of the diamond chain compound Cu$_{3}$(CO$_{3}$)$_{2}$(OH)$_{2}$\cite%
{CuCOOH1} can be nicely fitted by our TMRG results.

For the spin-$1/2$ frustrated Heisenberg diamond chains with AF couplings $%
J_{1}$, $J_{2}$ and $J_{3}$, the magnetization plateau at $m=1/6$ in the
ground state coincides with a perfect fixed sequence of the averaged local
magnetic moment such as $\{...,(S_{a},S_{a},S_{b}),...\}$ with $%
2S_{a}+S_{b}=1/2$, which might be described by trimerized states. On the
other hand, the static structure factor $S(q)$ shows peaks at wave vectors $%
q=0$, $\pi /3$ ($5\pi /3$), and $2\pi /3$ ($4\pi /3$) for different external
fields and different AF couplings. We note that the similar behavior of $S(q)
$ has been experimentally observed in diamond-typed compound Sr$_{3}$Cu$_{3}$%
(PO$_{4}$)$_{4}$ \cite{SrCu}. In addition, the DMRG results of the
zero-field static structure factor can be nicely fitted by a linear
superposition of six modes. It is observed that the six modes are closely
related to the low-lying excitations of the present case. At finite
temperatures, the magnetization $m(H)$, susceptibility $\chi (T)$ and
specific heat $C(T)$ demonstrate different behaviors at different AF
couplings, say, the magnetization plateau at $m=1/6$ is observed whose width
is found to be dependent on the couplings; the double peak structure is
observed for the susceptibility $\chi (T)$ and specific heat $C(T)$ as a
function of temperature, and the heights and positions of the peaks are
found to be dependent on the AF couplings.

For the spin-$1/2$ frustrated Heisenberg diamond chains with F couplings $%
J_{1}$, $J_{3}$ and AF coupling $J_{2}$, the magnetization plateau at $m=1/6$
in the ground state corresponds to a perfect fixed sequence of the averaged
local magnetic moment such as $\{...,(S_{a},S_{b},S_{b}),...\}$ with $%
S_{a}+2S_{b}=1/2$, which could be understood by trimerized states. The
static structure factor $S(q)$ shows peaks also at wave vectors $q=0$, $\pi
/3$ ($5\pi /3$), and $2\pi /3$ ($4\pi /3$) for different external fields and
different F couplings $J_{1}$, $J_{3}$ and AF coupling $J_{2}$, which is
expected to be experimentally observed in the related diamond-type compound.
In addition, the DMRG results of the zero-field static structure factor can
be nicely fitted by a linear superposition of six modes with the fitting
equations mentioned above. The six modes are closely related to the
low-lying excitations of the system. At finite temperatures, the
magnetization $m(H)$, susceptibility $\chi (T)$ and specific heat $C(T)$
demonstrate various behaviors for different couplings, namely, the
magnetization plateau at $m$ = $1/6$ is observed whose width is found to
depend on the couplings; the double-peak structure is also observed for the
susceptibility $\chi (T)$ and specific heat $C(T)$, and the heights and
positions of the peaks are found dependent on F couplings $J_{1}$, $J_{3}$
and AF coupling $J_{2}$.

For the spin-$1/2$ non-frustrated Heisenberg diamond chains with AF
couplings $J_{1}$, $J_{2}$ and F coupling $J_{3}$, the magnetization plateau
at $m=1/6$ in the ground state coincides with a perfect fixed sequence of
the averaged local magnetic moment such as $\{...,(S_{a},S_{b},S_{b}),...\}$
with $S_{a}+2S_{b}=1/2$, which could be understood by trimerized states. The
static structure factor $S(q)$ is observed to exhibit the peaks at wave
vectors $q=0$ and $2\pi /3$ ($4\pi /3$) for different external fields and
different AF couplings $J_{1}$, $J_{2}$ and F coupling $J_{3}$, which could
be experimentally detected in the related diamond-type compound. In
addition, it is found that the zero-field spin correlation function $\langle
S_{j}^{z}S_{0}^{z}\rangle $ is similar to that of the $S=1/2$ Heisenberg AF
chain. At finite temperatures, the magnetization $m(H)$, susceptibility $%
\chi (T)$ and specific heat $C(T)$ are found to reveal different behaviors
for different couplings, i.e., the magnetization plateau at $m=1/6$ is
obtained, whose width is found to depend on the couplings; the double-peak
structure is observed for the temperature dependence of the susceptibility $%
\chi (T)$ and specific heat $C(T)$, where the heights and positions of the
peaks depend on different AF couplings $J_{1}$, $J_{2}$ and F coupling $J_{3}
$.

The effect of the anisotropy of the AF and F interactions on the physical
properties of the non-frustrated Heisenberg diamond chain is also
investigated. For the case of the couplings satisfying $%
J_{1}:J_{2z}:J_{3}=1:2:-0.5$, when the anisotropic ratio $\gamma
_{2}=J_{2x}/J_{2z}=J_{2y}/J_{2z}\neq 1$, it is found that the width of the
plateau at $m=1/6$, the saturation field, and the susceptibility $\chi (T)$
show the same tendency, but quantitatively different, under the external
field $H$ along the $z$ and $x$ directions, while the specific heat $C(T)$
for $H$ along the $z$ direction coincides with that along the $x$ direction.
For the case of the couplings satisfying $J_{1}:J_{2}:J_{3z}=1:2:-0.5$, when
the anisotropic ratio $\gamma _{3}=J_{3x}/J_{3z}=J_{3y}/J_{3z}\neq 1$, it is
seen that the width of the plateau at $m=1/6$, the saturation field, and the
susceptibility $\chi (T)$ exhibit the opposite trends for $H$ along the $z$
and $x$ directions, while the specific heat $C(T)$ for $H$ along the $z$
direction also coincides with that along the $x$ direction.

For all the three cases, plateau states of $m=1/6$ are observed during the
magnetization, whose static structure factor $S(q)$ shows peaks at
wavevectors $q=0$, $2\pi /3$ and $4\pi /3$. But in absence of the magnetic
field, the static structure factor $S(q)$ in the ground state displays peaks
at $q=0$, $\pi /3$, $2\pi /3$, $\pi $, $4\pi /3$, and $5\pi /3$ for the
frustrated case with $J_{1}$, $J_{2}$, $J_{3}>0$; peaks at $q=0$, $\pi /3$, $%
\pi $, and $5\pi /3$ for the frustrated case with $J_{1}$, $J_{3}<0$, $%
J_{2}>0$; and a peak at $q=\pi $ for the non-frustrated case with $J_{1}$, $%
J_{2}>0$, $J_{3}<0$. In addition, the DMRG results of the zero-field static
structure factor can be nicely fitted by a linear superposition of six
modes, where the fitting equation is proposed. At finite temperatures, the
double-peak structure of the susceptibility and specific heat against
temperature can be obtained for all the three cases. It is found that the
susceptibility shows ferrimagnetic characteristics for the two frustrated
cases with some couplings, while no ferrimagnetic behaviors are observed for
the non-frustrated case.

The compound Cu$_{3}$(CO$_{3}$)$_{2}$(OH)$_{2}$ is regarded as a model
substance for the spin-$1/2$ Heisenberg diamond chain. The $1/3$
magnetization plateau and the two broad peaks both in the magnetic
susceptibility and the specific heat have been observed experimentally\cite%
{CuCOOH1}. Our TMRG calculations with $J_{1}:J_{2}:J_{3z}=1:1.9:-0.3$ and $%
J_{3x}/J_{3z}=J_{3y}/J_{3z}=1.7$ capture well the main characteristics of
the experimental susceptibility, specific heat and magnetization, indicating
that the compound Cu$_{3}$(CO$_{3}$)$_{2}$(OH)$_{2}$ may not be a spin
frustrated magnet\cite{GuSu3}.

\begin{appendix}

\section{LOW-LYING EXCITATIONS OF SPIN-1/2 Frustrated
HEISENBERG DIAMOND CHAINS}

In this Appendix, the low-lying excitations of
the spin-1/2 frustrated Heisenberg diamond chain are
investigated by means of the Jordan-Wigner (JW) transformation.
The Hamiltonian of the system reads
\begin{eqnarray}
\mathcal{H} &=&\sum\limits_{i=1}^{N}(J_{1}\mathbf{S}_{3i-2}\cdot \mathbf{S}%
_{3i-1}+J_{2}\mathbf{S}_{3i-1}\cdot \mathbf{S}_{3i}
+J_{3}\mathbf{S}_{3i-2}\cdot \mathbf{S}_{3i} \notag\\
&&+J_{3}\mathbf{S}_{3i-1}\cdot \mathbf{S}_{3i+1}
+J_{1}\mathbf{S}_{3i}\cdot \mathbf{S}_{3i+1})- H
\sum\limits_{j=1}^{3N}S_{j}^{z}, \label{Ham}
\end{eqnarray}
where $3N$ is the total number of spins in the diamond chain,
$J_{i}>0$ ($i=1,2,3$) represent the AF coupling while $J_{i}<0$ the
F interaction, and $H$ is the external magnetic field along the $z$
direction. In accordance with the spin configuration of the diamond
chain, we start from the Jordan-Wigner (JW) transformation with
spinless fermions
\begin{eqnarray}
S_{j}^{+}&=&a_{j}^{+}exp[i\pi\sum_{m=1}^{j-1}a_{m}^{+}a_{m}], \notag \\
S_{j}^{z}&=&a_{j}^{+}a_{j}-\frac{1}{2},
\end{eqnarray}
where $j=1,\cdots,3N$. Because the period of the present system is 3,
three kinds of fermions in moment space can be introduced through the
Fourier transformations
\begin{eqnarray}
a_{3i-2} &=&\frac{1}{\sqrt{N}}\sum\limits_{k}e^{ik(3i-2)}a_{1k},
\notag \\
a_{3i-1} &=&\frac{1}{\sqrt{N}}\sum\limits_{k}e^{ik(3i-1)}a_{2k},
\notag \\
a_{3i} &=&\frac{1}{\sqrt{N}}\sum\limits_{k}e^{ik(3i)}a_{3k}.
\end{eqnarray}
Ignoring the interactions between fermions, the Hamiltonian takes
the form of
\begin{eqnarray}
H&=&E_{0}+\sum\limits_{k}[(\omega _{1}a_{1k}^{+}a_{1k} + \omega
_{2}a_{2k}^{+}a_{2k}+ \omega _{3}a_{3k}^{+}a_{3k})\notag \\
&&+(\gamma_{1}a_{1k}a_{2k}^{+}+\gamma_{2}a_{2k}a_{3k}^{+}
+\gamma_{3}a_{3k}a_{1k}^{+}+h.c.)],
\end{eqnarray}
where $E_{0}= \frac{N}{4}(2J_{1}+J_{2}+2J_{3}-6H)$, $\omega
_{1}$=$-(J_{1}+J_{3})-H$, $\omega
_{2}$=$-\frac{1}{2}(J_{1}+J_{2}+J_{3})-H$,
$\omega_{3}$=$\omega_{2}$,
$\gamma_{1}$=$(J_{1}e^{ik}+J_{3}e^{-i2k})/2$,
$\gamma_{2}$=$(J_{2}e^{ik})/2$, and
$\gamma_{3}$=$(J_{3}e^{ik}+J_{1}e^{-ik})/2$.

Via the Bogoliubov transformation
\begin{eqnarray}
a_{1k} &=& u_{11}(k)\alpha_{1k}
+u_{12}(k)\alpha_{2k}+u_{13}(k)\alpha_{3k}, \notag \\
a_{2k} &=& u_{21}(k)\alpha_{1k}
+u_{22}(k)\alpha_{2k}+u_{23}(k)\alpha_{3k}, \notag \\
a_{3k} &=& u_{31}(k)\alpha_{1k}
+u_{32}(k)\alpha_{2k}+u_{33}(k)\alpha_{3k},
\end{eqnarray}
the Hamiltonian can be diagonalized as
\begin{equation}
H=E_{g}+\sum\limits_{k}\sum\limits_{i=1}^{3}\epsilon_{ik}\alpha
_{ik}^{+}\alpha _{ik}.
\end{equation}
The coefficients of the Bogoliubov transformation can be found
through equations of motion $i\hbar \dot{a}_{ik}=[a_{ik},H]$:
\begin{equation}
\left(
\begin{array}{lcr}
\omega _{1} & \gamma _{1} & \gamma _{3} \\
\gamma _{1}^{\ast} & \omega _{2} & \gamma _{2} \\
\gamma _{3}^{\ast} & \gamma_{2}^{\ast}& \omega _{3}
\end{array}
\right) \left(
\begin{array}{c}
u_{1i} \\
u_{2i} \\
u_{3i}
\end{array}%
\right) =\epsilon_{ik}\left(
\begin{array}{c}
u_{1i} \\
u_{2i} \\
u_{3i}
\end{array}\right).
\end{equation}
For a given $k$, the eigenvalues $\varepsilon_{ik}$ and eigenvectors
$(u_{1i}, u_{2i}, u_{3i})$ can be numerically calculated by the driver
$ZGEEV.f$ of the $LAPACK$, which is available on the
website\cite{LAPACK}. Figs.\ref{aaaexcita} show the zero-field
low-lying fermionic excitation $\varepsilon (k)$ for the frustrated
diamond chain with different AF coupling, while Figs.\ref{fafexcita}
present the zero-field low-lying fermionic excitation
$\varepsilon(k)$ for the frustrated diamond chain with $J_{1}$,
$J_{3}<0$ and $J_{2}>0$.

\end{appendix}

\acknowledgments

We are grateful to Prof. D. P. Arovas for useful communication. This work is
supported in part by the National Science Fund for Distinguished Young
Scholars of China (Grant No. 10625419), the National Science Foundation of
China (Grant Nos. 90403036, 20490210, 10247002), and by the MOST of China
(Grant No. 2006CB601102).


\begin{thebibliography}{99}
\bibitem{note1} We have checked that in the present situation, the static
structure factor $S(q)$ calculated from $\langle S_{j}^{z}S_{0}^{z}\rangle $
coincides with that from $\langle (S_{j}^{z}-\langle S_{j}^{z}\rangle
)(S_{0}^{z}-\langle S_{0}^{z}\rangle )\rangle $.
\end{thebibliography}
\end{document}